\documentclass[prd,twocolumn,showpacs,superscriptaddress]{revtex4}

\usepackage{amsfonts}
\usepackage{amsmath}
\usepackage{bm}
\usepackage{graphicx,epsf,epsfig,amssymb}
\usepackage[latin1]{inputenc}
\usepackage{latexsym}
\usepackage{rotating}

\newcommand{\be}{\begin{equation}}
\newcommand{\ee}{\end{equation}}
\newcommand{\ba}{\begin{eqnarray}} 
\newcommand{\ea}{\end{eqnarray}}

\begin{document}

\def\jnl@style{\it}
\def\aaref@jnl#1{{\jnl@style#1}}

\def\aaref@jnl#1{{\jnl@style#1}}

\def\aj{\aaref@jnl{AJ}}                   
\def\apj{\aaref@jnl{ApJ}}                 
\def\apjl{\aaref@jnl{ApJ}}                
\def\apjs{\aaref@jnl{ApJS}}               
\def\apss{\aaref@jnl{Ap\&SS}}             
\def\aap{\aaref@jnl{A\&A}}                
\def\aapr{\aaref@jnl{A\&A~Rev.}}          
\def\aaps{\aaref@jnl{A\&AS}}              
\def\mnras{\aaref@jnl{MNRAS}}             
\def\prd{\aaref@jnl{Phys.~Rev.~D}}        
\def\prl{\aaref@jnl{Phys.~Rev.~Lett.}}    
\def\qjras{\aaref@jnl{QJRAS}}             
\def\skytel{\aaref@jnl{S\&T}}             
\def\ssr{\aaref@jnl{Space~Sci.~Rev.}}     
\def\zap{\aaref@jnl{ZAp}}                 
\def\nat{\aaref@jnl{Nature}}              
\def\aplett{\aaref@jnl{Astrophys.~Lett.}} 
\def\apspr{\aaref@jnl{Astrophys.~Space~Phys.~Res.}} 
\def\physrep{\aaref@jnl{Phys.~Rep.}}      
\def\physscr{\aaref@jnl{Phys.~Scr}}       

\let\astap=\aap
\let\apjlett=\apjl
\let\apjsupp=\apjs
\let\applopt=\ao

\title[Accuracy of the Post-Newtonian Approximation]
{Accuracy of the Post-Newtonian Approximation: \\
 Optimal Asymptotic Expansion for Quasi-Circular, Extreme-Mass Ratio
 Inspirals} 




\author{Nicol\'as Yunes} \affiliation{Institute for Gravitational Physics and
  Geometry and Center for Gravitational Wave Physics, \\
  Physics Department, The Pennsylvania State University, University Park, PA 16802, USA}

\author{Emanuele Berti} \affiliation{Jet Propulsion Laboratory, California
  Institute of Technology, Pasadena, CA 91109, USA}

\date{\today}

\begin{abstract}
  
  We study the accuracy of the post-Newtonian (PN) approximation and
  its formal region of validity, by investigating its {\emph{optimal
      asymptotic expansion}} for the quasi-circular, adiabatic
  inspiral of a point particle into a Schwarzschild black hole. By
  comparing the PN expansion of the energy flux to numerical
  calculations in the perturbative Teukolsky formalism, we show that
  (i) the inclusion of higher multipoles is necessary to establish the
  accuracy of high-order PN terms, and (ii) the region of validity of
  PN theory is largest at relative ${\cal{O}}(1/c^{6})$ (3PN order).
  The latter result suggests that the series diverges beyond 3PN
  order, at least in the extreme mass-ratio limit, probably due to the
  appearance of logarithmic terms in the energy flux.  The study
  presented here is a first formal attempt to determine the
  region of validity of the PN approximation using asymptotic
  analysis. Therefore, it should serve as a template to perform similar
  studies on other systems, such as comparable-mass quasi-circular
  inspirals computed by high-accuracy numerical relativistic simulations.
  
\end{abstract}

\preprint{IGC-08/3-1}
\maketitle

\section{Introduction}

%


Before recent breakthroughs in numerical relativity (see
e.~g.~\cite{Pretorius:2007nq} for a review), the post-Newtonian (PN)
expansion had long been regarded as the best tool to predict the
evolution of compact binaries. In spite of these great numerical
advances, the PN approximation is still essential for the early
inspiral phase, and especially for generic systems with arbitrary
spins and eccentricities. Therefore, this approximation is still
indispensable in the construction of templates for gravitational-wave
detection through Earth- and space-based interferometers.

An accurate knowledge of the gravitational-wave phase is crucial for
interferometric detection of compact binaries. In turn, the phasing
accuracy crucially depends on the convergence (or divergence)
properties of the PN expansion~\cite{Cutler:1992tc}. For this reason,
the structure of the PN approximation has been extensively studied
with the following two goals: to determine the region of validity of
the series; and to improve its accuracy through resummation
techniques.  We shall not discuss the latter problem here, but we
refer the reader
to~\cite{Damour:1997ub,Damour:2000gg,Buonanno:1998gg,Buonanno:2000ef,Damour:2007yf,Porter:2005cu,Porter:2006cn,Porter:2007vk}.

Early studies of the accuracy of the PN approximation focused on head-on
collisions and on the quasi-circular inspiral of extreme-mass ratio (EMR)
compact binaries.  Simone {\emph{et al.~}}\cite{Simone:1995qu,Simone:1996db}
compared the relative ${\cal{O}}(1/c^4)$-accurate expansion of the energy flux
to numerical perturbative calculations. They found that the series converges
slowly for particles falling radially into black holes, remaining accurate for
$v/c \lesssim 0.3$, where $v$ is the orbital velocity and $c$ is the speed of
light. Building on previous
work~\cite{Poisson:1993vp,Poisson:1994yf,Tagoshi:1994sm,Tanaka:1997dj},
Poisson~\cite{Poisson:1995vs} compared the energy flux for quasi-circular EMR
inspirals computed with an ${\cal{O}}(1/c^{11})$-accurate PN expansion of black
hole perturbation equations to numerical results. Poisson found that the PN
series performs poorly for $v/c \gtrsim 0.2$, and that higher-order terms do
not necessarily increase the detection performance of the series, as measured
by the fitting-factor with numerical waveforms. Various authors argued that
the PN series should converge much faster for comparable-mass binaries
(see e.~g.~\cite{Simone:1996db,Blanchet:2002xy,Mora:2003wt}).

Previous studies of the accuracy of the PN expansion have been
necessarily limited by the lack of accurate (ideally exact) numerical
solutions of the non-linear Einstein equations, especially for the
comparable-mass case. As increasingly accurate numerical evolutions of
compact binaries become available, such convergence studies should be
revisited, taking into account the improved knowledge of the ``true''
numerical evolution of the system.  Until now, however, a systematic
asymptotic analysis of the accuracy of the PN series was lacking.  The
term ``asymptotic analysis'' does not here refer to the study of the
structure of the series at future or past null infinity. Instead, we
mean those techniques applicable to {\emph{asymptotic series}}, which
arise as approximate solutions to non-linear partial differential
equations.

In this paper, we perform an asymptotic analysis of the accuracy of
the PN approximation by investigating the quasi-circular inspiral of
EMR compact binaries.  Asymptotic methods assume nothing about the
convergence of the series, but only that it derives from the
approximate solution to a {\emph{consistent}} system of differential
equations (see e.~g.~\cite{Bender} for an introduction). In
particular, we shall here search for the {\emph{optimal asymptotic
    expansion}} of the PN series, i.~e. for the truncation order
beyond which the error in the series becomes larger than expected
(that is, larger than the next term in the series). The main goal of
this paper is to determine the approximate region of validity of the
PN approximation for for quasi-circular EMR systems as a function of
PN order.

The approximate region of validity is bounded by the region where the
true error in the PN series is comparable to the PN error estimate. By
``true error'' we here mean the difference between the PN estimate and
the ``exact solution'' (i.~e.~the numerical result), while the ``PN
error estimate'' is simply the next order term in the series. The
orbital velocity will serve as the independent variable that labels
this region, since this is a coordinate-invariant quantity, which for
EMR systems is related to the angular velocity via $\omega = v^3/M$,
with $M$ the total mass of the system.  The orbital velocity beyond
which these two errors become comparable marks the region outside
which one cannot neglect higher-order terms in the PN expansion. This
is simply because, for larger velocities, the next order terms in the
series are as large as the true error in the approximation.

Such an analysis requires we study the temporal evolution of the
approximate solution to the Einstein equations. This could be achieved
by investigating the PN expansion of several different quantities,
such as the energy flux or the metric perturbation. The specific
choice of PN-expanded quantity should not strongly affect the region
of validity estimates, since all such quantities are expanded
consistently to the same order. Here we choose to work with the energy
flux, which is also an observable and a coordinate-invariant quantity.

The consistency of this analysis hinges, of course, on the error contained in
the ``exact'' numerical solution. For the asymptotic analysis to succeed, this
numerical error must be smaller than the PN error estimate. For example, if we
investigate the region of validity of some PN quantity to relative
${\cal{O}}(1/c^4)$, the numerical error must be smaller than the terms of
relative ${\cal{O}}(1/c^5)$ in the series. The energy flux of EMR systems can
be accurately modeled through the Teukolsky perturbative formalism, which,
coupled to Green-function methods and spectral integrators, guarantees here a
numerical accuracy of ${\cal{O}}(10^{-6})$. We shall see that this numerical
accuracy suffices to perform an asymptotic analysis of the PN series to
relative ${\cal{O}}(1/c^{11})$ in the most interesting range of the particle's
orbital velocity.

\begin{figure}[htb]
\epsfig{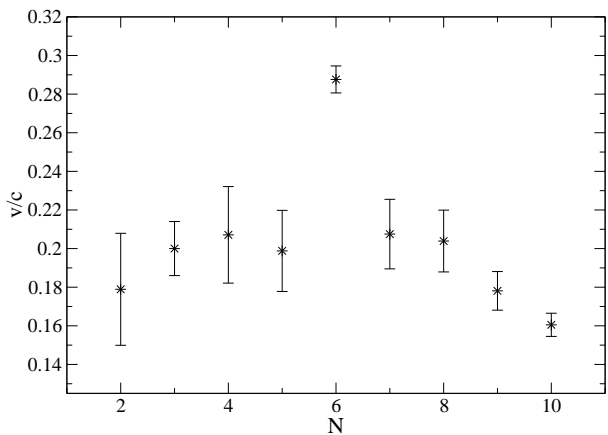} 
\caption{\label{intro:regions-final-PN-EMR} Edge of the region of validity for
  different PN orders.}
\end{figure}
The edge of the region of validity of the PN series for quasi-circular
EMR compact binaries as a function of PN order $N$ is given by
Fig.~\ref{intro:regions-final-PN-EMR}. These results verify and extend
those of Simone, {\emph{et al.}}~\cite{Simone:1995qu,Simone:1996db}.
The error bars in this figure symbolize the uncertainty inherent in the
definition of the edge of the region of validity of any asymptotic series,
which is defined more accurately in Sec.~\ref{asymptotics} below. For relative
${\cal{O}}(1/c^6)$ and smaller the PN solution seems to have a ``convergent''
character, since the region of validity either increases or remains roughly
constant with PN order.  On the other hand, for larger than relative ${\cal{O}}(1/c^6)$
there is a ``divergent'' behavior, reflected in the shrinking of the region
of validity with increasing PN order. As we shall see, this may be associated
with the appearance of logarithms at high orders in the PN approximation.
Similar conclusions were reached by Porter
\cite{Porter:2005cu,Porter:2006cn,Porter:2007vk}, when studying how to
increase the accuracy of the PN series through Chebyshev resummation.

A by-product of this analysis is the determination of the
{\emph{minimum}} number of multipoles needed in the numerical energy
flux to perform {\emph{any}} type of comparison with the PN approximation as a
function of velocity. These results are presented in
Table~\ref{table:lmin}, which provides an easy reference to determine
how many multipoles are needed to have an accuracy comparable to the
${\cal{O}}(1/c^N)$ PN approximation. For example, if one requires an
accuracy of relative ${\cal{O}}(1/c^6)$, then one need only include up
to $\ell = 2$ between $0.298<v/c<0.408$, up to $\ell = 3$ between
$0.131<v/c<0.298$ and up to $\ell = 4$ for $v/c < 0.131$.  Remarkably,
for large velocities one can usually simply look at the $\ell = 2$
multipole, while for small velocities one must include more and more
multipoles. Such a result is a consequence of individual multipolar
contributions being fairly velocity-independent in the large velocity
regime.

The analysis presented in this paper provides a general,
gauge-independent and systematic method to study the region of
validity of the PN approximation for any system.  We concentrate on
EMR binaries here for simplicity, but the method can be
straightforwardly extended to comparable-mass, spinning or eccentric
binaries.  This method is more systematic and general than that of
Simone, {\emph{et al.~}}\cite{Simone:1996db}, and it is similar in
spirit to the ``PN diagnostic''
scheme~\cite{Mora:2003wt,Mora:2002gf,Berti:2006bj,Berti:2007cd}. In
this scheme, however, comparisons with numerical simulations are
carried out considering time-independent, conserved quantities in a
quasi-equilibrium framework. We hope that the analysis presented here
provides a template for numerical relativity groups to test the
convergence (or divergence) of the PN approximation, which is of great
interest both to the PN and data analysis communities.

The remaining of this paper presents more details of our methods and results,
and it is organized as follows.  Section~\ref{asymptotics} summarizes the
basics of asymptotic analysis, formally defining optimal asymptotic expansions
and providing a pedagogical example of how to determine the region of validity
of an asymptotic series. Section~\ref{EMRIs} reviews the energy flux of EMR
compact binaries in quasi-circular orbits as obtained in perturbation theory
and PN theory. Section~\ref{reg-of-val} calculates the region of validity of
the PN approximation through asymptotic techniques. Section~\ref{mult-exp}
concentrates on the number of multipoles needed in the numerical energy flux
to test PN theory.  Section~\ref{conclusions} concludes and points to future
research.

\begin{widetext}
\begin{center}
\begin{table}[htb]
\begin{tabular}{c|c|c|c|c|c|c|c|c|c|c}
\hline
\hline
$\ell_{\rm min}$ & $N=2$ & $N=3$ & $N=4$ & $N=5$ & $N=6$ & $N=7$ & $N=8$ & $N=9$ & $N=10$ & $N=11$\\
\hline
2 & $0.408$ & $0.408$ & -       & $0.408$ & $0.408$ & $0.408$ & -       & $0.408$ & -       & -       \\
3 & -       & $0.101$ & $0.408$ & $0.298$ & $0.306$ & $0.402$ & $0.408$ & $0.379$ & $0.408$ & -       \\
4 & -       & -       & -       & $0.057$ & $0.131$ & $0.251$ & $0.303$ & $0.275$ & $0.322$ & $0.408$ \\
5 & -       & -       & -       & -       & -       & $0.039$ & $0.123$ & $0.144$ & $0.212$ & $0.349$ \\
6 & -       & -       & -       & -       & -       & -       & -       & -       & $0.067$ & $0.247$ \\
\hline
\hline
\end{tabular}
\caption{\label{table:lmin} Minimum velocity at which we need to include a
  minimum of $\ell_{\rm min}$ multipoles to get an accuracy corresponding to
  the $N$th PN term.}
\end{table}
\end{center}
\end{widetext}

In this work we follow the conventions of Misner, Thorne and Wheeler
\cite{Misner:1973cw}: the metric has signature $(-,+,+,+)$, and unless
otherwise specified we use {\emph{geometrical units}} where $G=c=1$.
The notation ${\cal{O}}(A)$ stands for terms of relative order $A$,
where $A$ is dimensionless.  Therefore, PN remainders are denoted as
${\cal{O}}(v/c)^q = {\cal{O}}(1/c^q) = {\cal{O}}(v^q)$, where $q$ is
an integer. When we say a quantity is of $p$th PN order we mean that
it is accurate to relative ${\cal{O}}(1/c^{2p})$.  The standard
asymptotic notation will also be used extensively in this paper, and
it is defined in the next section.

\section{Asymptotic series}
\label{asymptotics}

In this section we shall review the basic properties of asymptotic
series.  We refer the reader to~\cite{Bender,Kevorkian} for more
details. We begin by defining remainders and asymptotic series, and
continue with a description of their convergent and divergent
properties. We then define the concept of an optimal asymptotic
expansion, using which we can estimate the region of validity of an
asymptotic series. We conclude this section with a pedagogical example
of such a calculation applied to Bessel functions.

\subsection{Basic definitions and notation}

Consider some partial or ordinary, linear or non-linear differential
equation, whose solution is $f(x)$. Consider also a partial sum of the
form
\be\label{powers}
s_N (x) = \sum_{n=0}^{N} a_n \left(x - x_0\right)^n\,.
\ee
We here choose such a partial sum for simplicity, but in general there
could be a controlling factor, such as an exponential function, that
multiplies the partial sum.  Let us then define the remainder
$\epsilon_{(N)}(x)$ as
\be
\epsilon_{(N)}(x) \equiv f(x) - s_N(x)\,.
\label{remainder}
\ee

The limit as $N\to \infty$ of the partial sum, $s_{\infty}(x)$, is said to be
{\emph{asymptotic to}} the function $f(x)$ as $x \to x_0$, i.~e.~$f(x) \sim
s_{\infty}(x) \;\; {\textrm{as}} \;\; (x \to x_0)$, if and only if
\be
\epsilon_{(N)}(x)  \ll \left(x - x_0\right)^N\,, \qquad \left(x \to x_0\right)\,,
\label{asymptoticreq}
\ee
for all $N$. The symbol $\sim$ here is used exclusively to mean
``asymptotic to'' and is never intended to mean ``approximately,''
which shall be denoted via the symbol $\approx$. Similarly, the symbol
$\ll$ has a specific definition: if $f(x) \ll g(x)$ as $x \to x_0$, then
\be
\lim_{x \to x_0} \frac{f(x)}{g(x)} = 0\,.
\ee
Then, $f(x) \sim s_{\infty}(x)$ as $x \to x_0$ if and only if
\be\label{asymptdef}
\lim_{x \to x_0} \frac{f(x)}{s_{N}(x)} = 1\,.
\ee
There are more formal definitions of an asymptotic series, but this
one will suffice for the analysis of this paper. From this definition,
it follows that all Taylor series are asymptotic series as well.

Such a definition of an asymptotic series has a clear physical
meaning: a power series is asymptotic to some function if the
remainder after $N$ terms is much smaller than the last retained term
as $x \to x_0$. In this sense, the PN expansion of any quantity is an
asymptotic series to the exact solution of the Einstein equations as
$v \to 0$. This is so because one expects there to exist a velocity
region where the remainder of the PN expansion after $N$ terms is much
smaller than the last retained term in the series. Such a remainder,
of course, can only be computed once an exact or numerical solution is
known.

\subsection{Convergent and divergent series}

An important consequence of the above definition is that asymptotic series
need not be convergent. A series is convergent if and only if
\be
\label{convergent}
\lim_{N \to \infty} \epsilon_{(N)} = 0\,, 
\ee
for all $x$ inside a given radius of convergence $R$ around $x_0$,
i.~e.~for $|x - x_0| < R$. The radius of convergence can be computed
via the standard Cauchy ratio test:
\be
R = \lim_{n \to \infty} \frac{a_{n}}{a_{n+1}}\,,
\ee
where $a_n$ is the $n$th term in the series. 

The convergence requirement of Eq.~(\ref{convergent}) is much stronger
than the asymptotic requirement of Eq.~(\ref{asymptoticreq}). The main
difference is that in the asymptotic definition the remainder need not
go to zero as $N \to \infty$. Thus, asymptotic series can be
divergent. In fact, such series often approach the exact or numerical
solution much faster than any convergent series.  However, if one
insists on adding higher-order terms to the series for some fixed
value of $x$ (or if one pushes the approximation to $x \gg x_0$),
eventually the series will diverge.  Thus, a correct use of asymptotic
series forces us to truncate them before the answer deviates too much
from the exact solution.

The concept of convergence is said to be {\emph{absolute}}, because it
is an intrinsic property of the coefficients $a_n$ and it requires no
knowledge of the exact solution $f(x)$. On the other hand, the concept
of asymptoticity is said to be {\emph{relative}} because it requires
knowledge both of the coefficients and of the exact or numerical
solution. This is the reason why analyses of the asymptoticity of the
PN series were not possible before the recent numerical relativity
breakthroughs.

\subsection{Optimal asymptotic expansion}

Suppose that we only know a limited number of terms in a (possibly divergent)
series. We want to determine the optimal number to include in the partial sum
to get an answer as close as possible to the exact solution, i.~e. the
so-called {\emph{optimal asymptotic expansion}}. In asymptotic language, we
are looking for the partial sum that minimizes the remainder in the
approximation.

More precisely, the procedure is as follows: (i) Choose some fixed value of
$|x - x_0|$, so that the series becomes a sum of $N$ numerical coefficients.
In PN theory, these coefficients will in general be functions of the symmetric
mass ratio $\eta\equiv m_1m_2/(m_1+m_2)^2$, of the spins of the binary members
and of the eccentricity parameters; (ii) For this fixed $|x-x_0|$, search over
the individual coefficients in the series.  Typically these terms initially
decrease in magnitude, but eventually diverge; (iii) Let the first minimum in
the sequence occur when $n=M$ such that $M < N$; (iv) The partial sum of all
terms in the series up to (but not including) the $M$th term is the optimal
asymptotic expansion. The $M$th term is also an approximate measure of the
error in the optimal asymptotic expansion, because it is asymptotic to the
$M$th remainder as $x \to x_0$. Thus, the optimal asymptotic expansion
minimizes the remainder, since if we kept on adding more terms the remainder
would diverge.

This method works nicely when we have a large number of terms in the series.
Unfortunately, it does not work very well in PN theory, where usually only a
few terms are known. Moreover, this method depends on the chosen value of
$|x-x_0|$. In PN theory, it is precisely this value of $|x - x_0|$ that we
wish to determine. Nonetheless, we can adapt the method to find the region of
validity of any PN series, as we show below.

\subsection{Region of validity}

Let us invert the premise and look for the {\emph{region of validity}} as a
function of the number of terms kept in the series. This region can be defined
as the region inside which the remainder $\epsilon_{(N)}$ and the $(N+1)$th
term in the series are of the same order. If the series is asymptotic to the
exact solution, we then expect $s_{N}$ to be the best approximation to $f(x)$
for all $x \leq \bar{x}$ with
\be
{\cal{O}}\left[\epsilon_{(N)}(\bar{x})\right] = {\cal{O}}
  \left(\bar{x} - x_0\right)^{N+1}\,. 
\label{region-of-validity}
\ee

The convergent or divergent character of the asymptotic series
determines how $\bar{x}$ behaves as a function of the number of terms
kept in the series. If the series is convergent, like a Taylor
expansion, then we expect the accuracy of the series to increase as
more terms are added.  This is so provided $\bar{x}$ is inside the
radius of convergence, but as already discussed, this is always the
case because Eq.~\eqref{convergent} is stronger than
Eq.~\eqref{asymptoticreq}.  Therefore, it follows that $\bar{x}$ tends
to increase, and in fact approach the radius of convergence, as the
number of terms kept in the series increases. On the other hand, if
the series is divergent, then the opposite is true, namely as more
terms are kept in the series, $\bar{x}$ decreases.  This is so because
higher-order terms tend to diverge faster than lower-order ones.

The order symbol encodes here a certain arbitrariness rooted in
asymptotic analysis. In other words, one must decide a priori how
different the right and left-hand side of
Eq.~\eqref{region-of-validity} should be before equality is declared.
This ambiguity means that there is not a precise value for the region
of validity of an asymptotic approximation. Instead, this concept is
ambiguous up to the order symbol. In spite of this ambiguity, we can
still arrive at important qualitative conclusions through such an
analysis, provided we quantify the ambiguity with error bars, as done
in later sections.

\subsection{A pedagogical example:  The modified Bessel function}

Before turning to an analysis of the PN series, in this section we
shall clarify the definitions of previous sections with a relatively
simple pedagogical example. Consider then the following differential
equation
\be
x^2 \frac{d^2 y}{dx^2} + x \frac{d y}{dx} - \left(x^2 + 25\right)
y = 0\,,
\label{Bessel}
\ee
One recognizes Eq.~\eqref{Bessel} as the modified Bessel equation with index
$\nu = 5$ (see e.~g.~\cite{abramowitz+stegun}). The modified Bessel equation
does not have a known closed-form solution valid everywhere in its domain, and
the numerical solution of this equation is called a modified Bessel function:
$y(x) = I_{5}(x)$.  The asymptotic expansion of the modified Bessel function
as $x \to \infty$, say $I_{5}^{(N)}$, can be found at any given order $N$.
For $N=12$, for example, we have:
\begin{widetext}
\ba
I_{5}^{(12)} &=& \frac{e^x}{\sqrt{2 \pi x}} 
\left[ 1 - \frac{99}{8}\frac{1}{x} + \frac{9009}{128}\frac{1}{x^2} 
- \frac{225225}{1024}\frac{1}{x^3}   +  \frac{11486475}{32768}\frac{1}{x^4} 
-\frac{43648605}{262144}\frac{1}{x^5} 
-\frac{305540235}{4194304}\frac{1}{x^6}  
- \frac{3011753745}{33554432}\frac{1}{x^7} 
\right. 
\nonumber \\
&-& \left.
\frac{376469218125}{2147483648}\frac{1}{x^8}  
-\frac{7905853580625}{17179869184}\frac{1}{x^9}  
-  
\frac{412685556908625}{274877906944}\frac{1}{x^{10}}  
-\frac{12793252264167375}{2199023255552}\frac{1}{x^{11}} 
\right. 
\nonumber \\
&-& \left.
\frac{1829435073775934625}{70368744177664}\frac{1}{x^{12}}  
+{\cal{O}}\left(1/x\right)^{13}\right]\,. 
\label{exp-bessel}
\ea
\end{widetext}
\begin{figure}[htb]
\epsfig{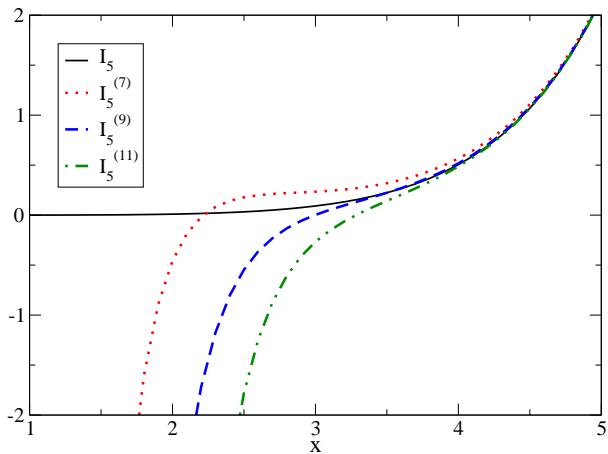}
\caption{\label{all-bessels} Plot of the modified Bessel function (solid)
  and its asymptotic expansion of order $N=7$ (dotted), $N=9$ (dashed)
  and $N=11$ (dot-dot-dashed).}
\end{figure}
Figure~\ref{all-bessels} shows the exact numerical solution to the
modified Bessel function (solid) and its $N=7$ (dotted), $N=9$
(dashed) and $N=11$ (dot-dot-dashed) asymptotic expansions.  The
figure clearly shows two important features of this asymptotic series:
(i) The series can be extremely accurate quite far from its singular
point of expansion $x = \infty$. In principle, there is no {\emph{a priori}} reason
to believe that any of these approximations will be accurate in the
domain plotted, since $x=5$ is far from $x=\infty$.
Nonetheless, all expansions plotted are
quite close to the numerical answer, even up to $x \approx 3$; (ii)
The higher the order $N$ of the asymptotic expansion, the smaller its
region of validity.  Observe that the $N=11$ ($N=7$) expansion roughly
deviates from the numerical answer when $x \approx 3$ ($x\approx 2$).
As already discussed, this behavior is usually associated with
divergent asymptotic series near the edge of their region of validity.

As a check of this statement, let us now determine the region of validity of
this asymptotic expansion. As explained earlier, this region will depend on
$N$, the order of the asymptotic expansion.  From Eq.~(\ref{remainder}) we see
that the remainder is given by $\epsilon_{(N)} = I_5 - I_5^{(N)}$. Noting that
the $N$th order term in the asymptotic series is simply $I_5^{(N)} -
I_5^{(N-1)}$, the region of validity is defined by the relation
[Eq.~\eqref{region-of-validity}]
\be
{\cal{O}}(I_5 - I_5^{(N)}) = {\cal{O}}(I_5^{(N+1)} - I_5^{(N)})\,.
\ee
Here it suffices to search for the intersection between these two curves.

\begin{figure}[htb]
\epsfig{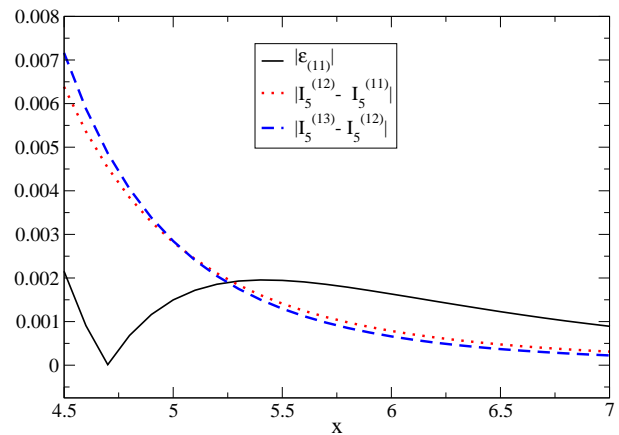}
\caption{\label{region-11} Plot of the absolute value of the remainder of the
  $N=11$ asymptotic expansion of the modified Bessel function (solid), the
  $N=12$ term (dotted) and the $N=13$ term (dashed). The kink in the solid
  line is due to the use of the absolute value operator.}
\end{figure}
Figure~\ref{region-11} plots the absolute value of the remainder of
the $N=11$ expansion (solid), the $N=12$ term (dotted) and the $N=13$
term (dashed).  For $x \gtrsim 5.25$ the exact error in the $N=11$
expansion relative to the numerical answer is larger than the terms
neglected in the approximation, while for $x \lesssim 5.25$ the
opposite is true.  Therefore, the intersection of the solid and dotted
curves defines {\emph{approximately}} the edge of the region of
validity for the $N=11$ partial sum: $\bar{x} \approx 5.25$.

\begin{table}
\begin{tabular}{c|c|c|c}
\hline
\hline
$n$  & $\bar{x}$ & $\delta\bar{x}$ & $\delta I_5^{(N)}$ [$\%$] \\
\hline
$11$ & $5.27$ & $0.22$ & $0.06$ \\
$10$ & $4.70$ & $0.26$ & $0.34$ \\
$9$ & $4.11$ & $0.03$ & $2.16$ \\
$8$ & $3.51$ & $0.04$ & $17.7$ \\
\hline\hline
\end{tabular}
\caption{\label{table1} Approximate edge of the region
  of validity ($x = \bar{x}$) for different asymptotic expansions of
  the modified Bessel function. We also present an approximate measure
  of the error in $\bar{x}$, as well as the fractional relative error
  in $I_5^{(N)}$, that is $\delta I_5^{(N)} = 100 (I_5 -
  I_5^{(N)})/I_5$, evaluated at $\bar{x}$.}
\end{table}

This analysis can be repeated for any $N$ and the results are
presented in Table~\ref{table1}, which reproduces and extends 
results in Table $3.1$ of~\cite{Bender}. The error bars presented in
the third column are given by the difference between $\bar{x}$ and the
intersection of the $(N+2)$-term with $N$th-order remainder. They
serve as a reminder that all quantities are asymptotic in nature, and
thus, can only be interpreted in an {\emph{approximate}} sense.

Table~\ref{table1} presents several features that are of interest.
First, observe that the region of validity decreases as
the order of the expansion increases (the region of validity is
$\{\bar{x},\infty\}$, so if $\bar{x}$ increases, the region of  
validity decreases).  As explained earlier, this is an important
feature of divergent asymptotic expansions, due to higher-order terms
diverging sooner than lower-order ones as we approach the edge of
the region of validity. Second, observe that higher-order partial sums
are much more accurate than low-order ones, when evaluated at their
respective edges of validity. For example, the $N=11$ expansion has an
error of only $0.06 \%$ relative to the exact numerical answer when
evaluated at $x \approx 5.3$. This is a sensible feature of optimal
asymptotic expansions, i.~e.~higher-order approximations should be
more accurate than lower-order ones.

The region of validity of an asymptotic expansion should be interpreted with
caution. In fact, the asymptotic expansion of the Bessel function could be
used outside its region of validity, as defined here. The risk of using the
$N$th-order approximation beyond $\bar{x} - \delta\bar{x}$ is that of making
errors {\em larger} than those supposedly contained in the approximation.  For
example, the $N=11$ expansion of the Bessel function produces a fractional
error of $(0.2,3.5,36.5)\%$ relative to the numerical answer when evaluated at
$x=(4.5,4.0,3.5)$, which is always larger than the next-order term in the
approximation. Therefore, beyond the region of validity as defined here,
neglecting the next order term in the approximation leads to larger and larger
errors relative to the exact numerical solution.

\section{Extreme-mass ratio inspirals}
\label{EMRIs}

In this section we shall apply the previously-discussed asymptotic
tools to the inspiral of a small compact object around a supermassive
black hole (an EMR system).  We shall thus consider a black hole of
mass $m_1 = M$ and a much smaller object of mass $m_2 = \mu\ll M$. We
shall further focus on the adiabatic quasicircular inspiral phase,
where the radiation-reaction timescale is much larger than the orbital
period.

Such a system is an excellent testbed for the methods discussed above.  The PN
approximation of several quantities is known to very high order in the EMR
limit. Furthermore, we can numerically compute the energy flux with
perturbative techniques that are very accurate for all velocities.

Before turning to the study of the region of validity of the PN approximation,
in the following sections we briefly review the derivation and accuracy of the
numerical solution in black hole perturbation theory and the analytical
structure of the PN expansion of the flux. We also present a simple graphical
comparison of the PN results with the numerical solution.

\subsection{Numerical calculation of the energy flux}

An exact solution is necessary if we want to use asymptotic analysis
to determine the region of validity of the PN approximation in the
extreme mass-ratio limit (henceforth denoted PN-EMR approximation).
This ``exact'' solution can be found numerically through the use of
black hole perturbation theory. Let us then express the metric as a
background (in our case, the Schwarzschild metric) plus a perturbation:
\be
g_{\alpha
  \beta} = g_{\alpha \beta}^{\rm{(0)}} + h_{\alpha \beta},
\ee
with the perturbation assumed to be small, i.~e. $|h_{\alpha \beta}| \ll
|g_{\alpha \beta}^{\rm{(0)}}|$.  The Einstein equations are then linearized in
$h_{\alpha \beta}$ and rewritten in terms of the perturbed Weyl Scalar $\delta
\Psi_4(t,r,\theta,\phi)$.  In turn, the perturbed Weyl scalar can be
decomposed into multipolar components.  Working in the frequency domain and
using spin-weighted spherical harmonics $_{-2}Y_{\ell m}(\theta,\phi)$, these
components are
\be 
\Psi_{\ell m}(\omega,r)=\frac{1}{2\pi}\int
d\Omega \;dt \; e^{i\omega t}~
_{-2}Y_{\ell m}^*(\theta,\phi)[r^4 \delta \Psi_4(t,r,\theta,\phi)]\,, 
\ee
The coefficients of the harmonic decomposition obey an inhomogeneous
differential equation, the Bardeen-Press-Teukolsky equation, where the source
term is the harmonic decomposition of the stress-energy tensor of the orbiting
particle. Due to the symmetries of the source term, the multipolar components
satisfy the symmetry relation
\be
\Psi_{\ell m}^*(r,\omega)=(-1)^\ell \Psi_{\ell -m}^*(r,-\omega)\,.
\ee
The Bardeen-Press-Teukolsky equation can be solved in the adiabatic
approximation with Green-function methods in the frequency-domain
\cite{Cutler:1993vq,Cutler:1994pb}. Our numerical code has been
described in detail in
\cite{Gualtieri:2001cm,Pons:2001xs,Berti:2002ry,Berti:thesis}. 

The perturbed Weyl scalar can be used to reconstruct the fluxes of
energy and momentum at infinity. This is so because these fluxes are
related to $h_{\alpha \beta}$ at infinity, which can be computed from
convolution integrals of $\Psi_{\ell m}(\omega,r)$ and the
(Fourier-decomposed) spherical-harmonic components of the
stress-energy tensor. For circular orbits, the multipolar components
of the energy flux $F_{\ell m}\equiv \dot E_{\ell m}$ (where $|m|\leq
\ell$) satisfy the symmetry property
\be
F_{\ell-m}=F_{\ell m}\,,
\ee
so we can limit consideration to multipoles with $m>0$. Therefore, when we
refer to ``the relative contribution of the $(l,m)$ multipole to the total
flux'' we are really considering {\em twice} $F_{\ell m}$. The angular
momentum flux is related to the energy flux through the particle's orbital
frequency $\Omega$, related to the orbital velocity via $v=(M\omega)^{1/3}$.
We have computed the energy flux including multipoles up to $\ell_{\rm max}=8$
for a thousand equispaced velocities in the range $v\in[10^{-2},v_{\rm
  ISCO}]$, where the orbital velocity at the ISCO $v_{\rm ISCO}=6^{-1/2}\simeq
0.408$.

The Teukolsky formalism has inherent errors, but it possesses some
distinct advantages over the PN expansion. The principal advantage is
that no restriction on the velocities is necessary: the numerical
solution is valid to all orders in $v$.  For this reason the numerical
solution is not only compatible with the PN-EMR approach, but it is
also very convenient to study the latter's region of validity.
Formally speaking, the PN-EMR approximation is a bivariate expansion
in two independent parameters: $v \ll 1$, such that the
gravitational field is weak and all bodies move slowly (PN
approximation); and $\mu/M \ll 1$, that enforces the EMR limit.
Therefore, there are two independent uncontrolled remainders of
relative ${\cal{O}}(N,M)$, which stands for errors of relative
${\cal{O}}(1/c)^N$ and ${\cal{O}}(\mu/M)^M$.  The perturbative
approach is valid to relative ${\cal{O}}(\infty,0)$, whereas the
PN-EMR expansion (discussed below) is only available up to relative
${\cal{O}}(11,0)$.

The asymptotic tools discussed in this paper hinge on controlling the
numerical error inherent in the ``exact'' solution. These errors cannot
always be modeled analytically (see \cite{Cutler:1993vq,Cutler:1994pb}
for a detailed discussion). When we consider the energy flux for a circular,
adiabatic inspiral, the dominant errors can be roughly divided in two groups:
\begin{description}
\item[(1) Truncation errors,] due to neglecting high-$\ell$ multipoles in the
  multipolar decomposition of $F_{\ell m}$;
\item[(2) Discretization errors,] due to solving numerically the inhomogeneous
  Teukolsky equation.
\end{description}

Truncation errors tend to increase with orbital velocity, and to a certain
extent they can be modeled analytically.  Poisson~\cite{Poisson:1993vp} has
shown that, for a given $\ell$, the luminosity is dominated by modes with even
$\ell+m$, while the power radiated by modes with odd $\ell+m$ is suppressed by
roughly a factor $v^2$. Since the total power radiated in the $\ell$th
multipole scales like
\be\label{truncation}
F_{\ell} \propto v^{(2\ell - 4)}\,, 
\ee
the error on the flux due to truncating at some $\ell=\ell_0$ can be
approximated by $\delta F_{\ell_0} \approx v^{2 \ell_0 -2}$.
Equation~(\ref{truncation}) provides a simple rule of thumb to
determine the number of multipoles required to achieve a given
fractional truncation accuracy $\epsilon_{(\ell)}$: we must include
multipolar components up to $\ell_{\rm max}$, where $v^{(2\ell_{\rm
    max}-4)}\sim \epsilon_{(\ell)}$ \cite{Cutler:1993vq}.  According
to this rule of thumb, including up to $\ell_{\rm max}=8$ (as we do,
unless otherwise stated) yields a truncation accuracy {\em better}
than $\epsilon_{(\ell)} \approx 10^{-6}$ up to $v\simeq 0.316$, while
at the ISCO we have an accuracy better than $\epsilon_{(\ell)}\approx
2\times 10^{-5}$.

Discretization error is kept small by use of adaptive ordinary
differential equation integrators to solve the Bardeen-Press-Teukolsky
equation, and Gauss-Legendre spectral methods to compute convolution
integrals~\cite{Press:1992nr}. We investigated this error by
experimenting with tolerance parameters in the integrators and
increasing the number of points in the spectral methods. We found that
discretization error at low velocities is very sensitive to other
details of the code, such as the accuracy in the numerical inversion
yielding $r(r_*)$, where $r$ is the Schwarzschild radius and $r_*$ is
the tortoise coordinate. At small velocities and large radii we obtain
agreement with Ref.~\cite{Cutler:1994pb} to a six-digit level, so we
estimate the discretization error to be roughly of
${\cal{O}}(10^{-6})$ (and possibly smaller at larger velocities). This
is confirmed by the low-velocity, low-amplitude region of
Fig.~\ref{remainder-PN-EMR} below, where oscillations due to
discretization error appear only for $v\lesssim 10^{-2}$. These
oscillations are in fact of ${\cal{O}}(10^{-7})$ relative to the
dominant Newtonian flux.

Before proceeding to the PN expansion of the energy flux in the EMR
limit, we must mention that both the perturbative calculation and the
PN expansion neglect absorption of radiation by the black hole.
Poisson and Sasaki~\cite{Poisson:1994yf} have shown that this
absorption is negligible with respect to the energy carried off to
infinity, being suppressed by a factor of $v^8$. Although these terms
might modify the energy flux ever so slightly, both the numerical and
PN flux considered here consistently neglect them. Therefore, these
terms cannot affect any conclusions regarding the region of validity
of the PN approximation.

\subsection{PN expansion of the energy flux}

The PN approximation is an expansion in small velocities $v \ll 1$,
but to model EMR systems we also expand in $\mu/M \ll 1$.  In this
PN-EMR approximation the energy flux is given
by~\cite{Tanaka:1997dj,Poisson:1995vs}
\be
F^{(N)} = F_{\mathrm{Newt}} 
\left[\sum_{k=0}^{N} \left(a_{k}+b_{k} \ln{v}\right) v^{k}\right]\,,
\label{E-EMRI-PN}
\ee
where $N$ labels the $N$th-order partial sum or PN approximant (known
up to $N=11$ in the PN-EMR approximation),
$a_1=b_1=b_2=b_3=b_4=b_5=b_7=0$, and the remaining coefficients can be
found in Eq.~(3.1) of Ref.~\cite{Tanaka:1997dj}.  As explained in the
Appendix, the logarithms do not affect the applicability of asymptotic
methods in the velocity regions of interest. Here $v$ is the orbital
velocity, and the Newtonian flux is given by
\be
F_{\mathrm{Newt}} = - \frac{32}{5} \mu^2 v^{10} M^2\,.
\ee

\begin{figure}[htb]
\epsfig{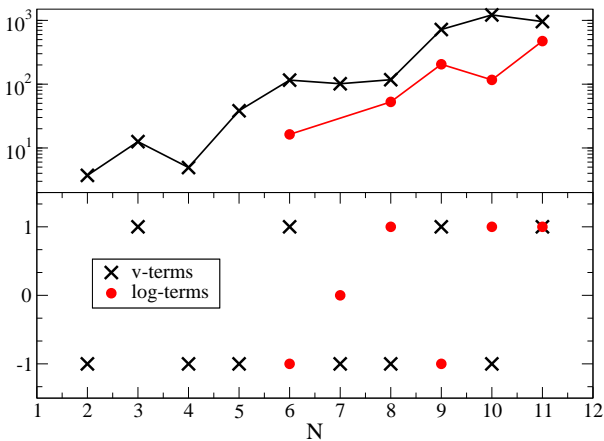}
\caption{\label{PN-order-3} Top panel: plot of the absolute magnitude of the
  coefficients of the flux as a function of PN order in a log-linear plot.
  Bottom panel: plot of the sign of the coefficients of the flux as a function
  of PN order. Crosses and circles denote polynomial ($a_k$) and logarithmic
  ($b_k$) terms, respectively.}
\end{figure}

The structure of the PN-EMR series is worth discussing further.  The series is
composed of two distinct types of terms: those that depend only on a certain
power of the velocity, and those that include a logarithmic dependence. Since
$b_6\neq 0$, the logarithmic terms appear first at ${\cal{O}}(1/c)^6$.
Furthermore, as shown graphically in Fig.~\ref{PN-order-3}, the sequence of
coefficients $a_{k}$ or $b_{k}$ does not present a clear alternating pattern.
Finally, the absolute magnitude of the coefficients in the series is not of
order unity, as expected of a convergent series. Instead, their magnitude
grows drastically with increasing order. Such a pathological behavior is not
present in the comparable-mass limit of the PN expansion (see
e.~g.~\cite{Blanchet:2002xy,Mora:2003wt} for a discussion).

\subsection{Comparison of PN-EMR and numerics}

\begin{figure}[htb]
\epsfig{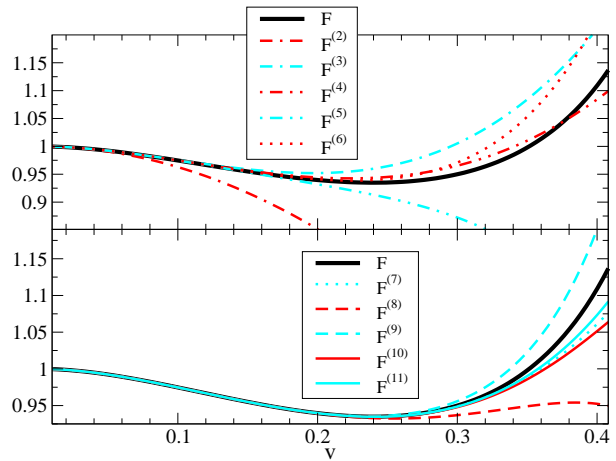}
\caption{\label{all-PN-EMRI} Plot of the total energy flux $F$ computed
  numerically (solid) and at different PN orders in the PN-EMR approximation.
  Odd orders are plotted in a lighter gray, while even orders are plotted in
  darker gray.  Line styles are as follows: $F^{(2)}$, $F^{(3)}$ is
  dash-dash-dot; $F^{(4)}$, $F^{(5)}$ is dash-dot-dot; $F^{(6)}$, $F^{(7)}$ is
  dotted; $F^{(8)}$, $F^{(9)}$ is dashed; $F^{(10)}$, $F^{(11)}$ is solid. To
  avoid cluttering, the top panel shows $F^{(2)}$ up to $F^{(6)}$ and the bottom
  panel shows $F^{(7)}$ up to $F^{(11)}$.}
\end{figure}

The energy flux computed both numerically and within the PN-EMR framework is
plotted in Fig.~\ref{all-PN-EMRI} as a function of orbital velocity. Here and
below, we shall present error bars in all figures only when these errors are
visible (e.~g.~if the $y$-axis scale is of ${\cal{O}}(10^{-1})$, the numerical
error won't be presented, since error bars of ${\cal{O}}(10^{-6})$ would not
be visible).  Figure~\ref{all-PN-EMRI} is very similar to Fig.~$1$
of~\cite{Poisson:1995vs}, except that here the numerical result is computed
including terms up to $\ell_{\rm max}=8$ and we show all terms up to 5.5PN
order in the flux.

For low velocities ($v \lesssim 0.2$) all PN approximants (except
for the 1PN curve) agree well with the numerical result, while in the
high-velocity region ($v \gtrsim 0.2$) the lower-order PN
approximants deviate from the numerical answer. Such a behavior is
reminiscent of that first discussed by Poisson~\cite{Poisson:1995vs}. Typically, and
roughly speaking, the partial sums seem to approach the numerical
result as we increase the PN order. The question we shall answer in
the next section is whether this approach occurs at the expected rate
for the given order.

\begin{figure}[htb]
\epsfig{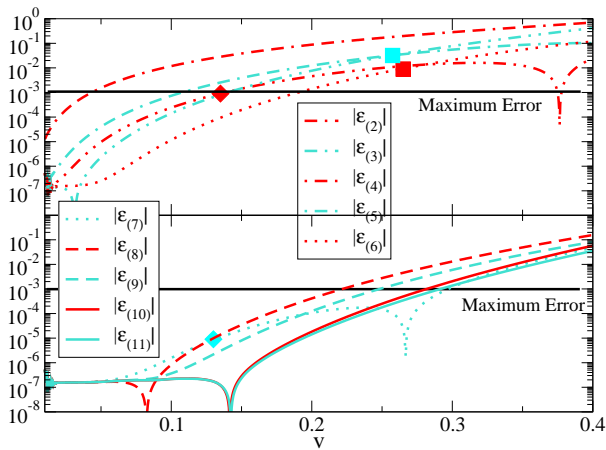}
\caption{\label{remainder-PN-EMR} Moduli of the remainders
  $|F^{(N)}-F|$ as a function of velocity. Light gray curves
  correspond to odd PN orders, while dark gray curves correspond to
  even orders. Line styles are the same as in Fig.~\ref{all-PN-EMRI}.
  The light gray (dark gray) squares mark the intersection of
  $\epsilon_{(4)}$ and $\epsilon_{(6)}$ ($\epsilon_{(3)}$ and
  $\epsilon_{(5)}$). The light gray (dark gray) diamonds mark the
  intersection of $\epsilon_{(4)}$ and $\epsilon_{(5)}$
  ($\epsilon_{(7)}$ and $\epsilon_{(8)}$). Numerical discretization
  errors at low velocities are of order $10^{-6}$ (see Section IIIA),
  therefore remainders smaller than about $10^{-6}$ have no physical
  meaning.}
\end{figure}
The accuracy of the PN approximants is better seen in
Fig.~\ref{remainder-PN-EMR}, where we plot the moduli of the
remainders $|F^{(N)}-F|$. Light and dark gray curves correspond to odd
and even PN orders respectively. We plot here only the region
$v<0.014$, since for smaller velocities the remainders are smaller
than ${\cal{O}}(10^{-6})$, and hence contaminated by numerical error.
Observe that the dominant even-order remainders decrease monotonically
with order until $\epsilon_{(4)}$ and $\epsilon_{(6)}$ cross at $v
\lesssim 0.257$ (dark gray square).  Similarly, the dominant odd-order
remainders also decrease monotonically until $\epsilon_{(3)}$ and
$\epsilon_{(5)}$ cross at $v \lesssim 0.265$ (light gray square). This
behavior is ``convergent'' in nature, but the region of convergence
seems to be larger when odd or even terms are considered separately.
Such a result derives perhaps from the fact that even and odd terms
are physically different, coming respectively from a sum over
multipoles moments and integrals of these over the entire propagation
history of the waves.

When all orders are considered simultaneously, the PN approximation
still presents a convergent-like behavior, but in a reduced region.
This reduction is due to the crossing of $\epsilon_{(4)}$ and
$\epsilon_{(5)}$ at $v \approx 0.135$ (dark gray diamond) and 
the crossing of $\epsilon_{(7)}$ and $\epsilon_{(8)}$ at $v \approx
0.125$ (light gray diamond).  Note also that the ``region of
convergence'' should not be confused with the ``region of validity.''
By the former we mean the region where the $(N+1)$th-order remainder
is smaller than the $N$th-order one. The latter is the region where
the asymptotic expansion is valid (the remainder is of the same order
as the next-order term), and it will be discussed in more detail
by asymptotic techniques in section \ref{reg-of-val} below.

In the large velocity region of Fig.~\ref{remainder-PN-EMR}, the
convergent-like behavior is replaced by a divergent one.  For example,
$\epsilon_{(4)}$ becomes more accurate than $\epsilon_{(5)}$ for $v >
0.135$.  Similarly, $\epsilon_{(7)}$ seems more accurate than
$\epsilon_{(8)}$ for $v > 0.125$ (see also Table~\ref{table0} in the
next section), and in fact $\epsilon_{(7)}$ is strikingly accurate
for $0.2<v<0.35$ --much better than $\epsilon_{(8)}$ and
$\epsilon_{(9)}$. This suggests that the region of validity of
lower-order approximations may be actually larger than that of
higher-order ones.  In other words, the region of validity seems to
{\emph{shrink}} with increasing order. Asymptotic techniques are ideal
to study the edge of the region of validity for high-velocities.

\begin{figure}
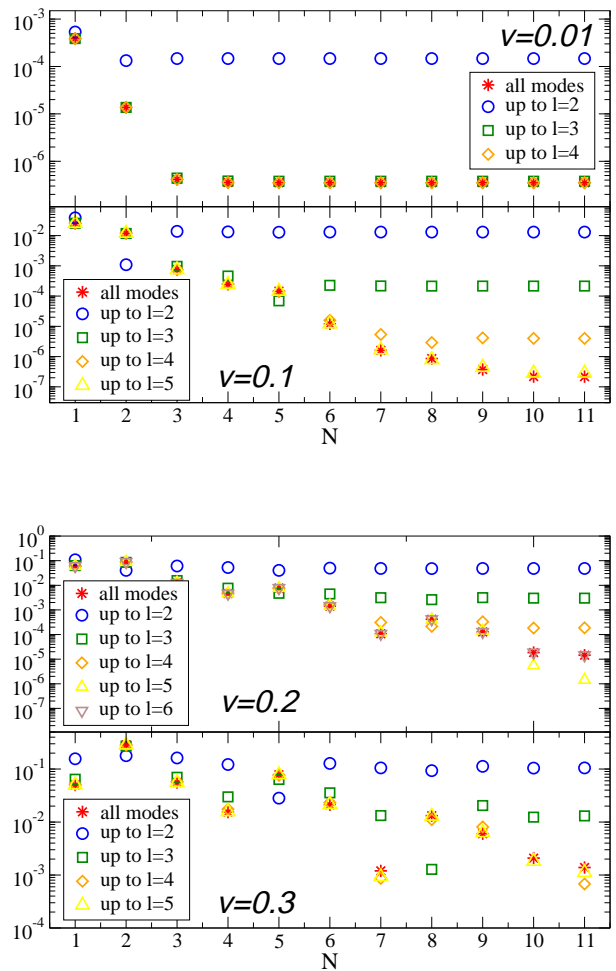

\epsfig{file=remainders-fix-v1.eps,width=8cm,angle=0,clip=true}
\\ \vspace{1cm}
\epsfig{file=remainders-fix-v2.eps,width=8cm,angle=0,clip=true}
\caption{\label{remainders-fix-v} Plot of the absolute value of the remainder
  of the energy flux at different values of $v$ versus PN order in a
  log-linear plot.  We present results obtained when summing
  multipoles up to $\ell = 2$ (circles), $\ell =3$ (squares), $\ell =
  4$ (diamonds), $\ell = 5$ (triangles up), $\ell = 6$ mode (triangles
  down) and $\ell = 8$ (stars).}
\end{figure}

The convergent/divergent transition is hard to see in
Fig.~\ref{remainder-PN-EMR}, so for clarity
Fig.~\ref{remainders-fix-v} plots the remainder as a function of PN
order $N$ for four selected values of $v$. Until now, we have always
included multipoles up to $\ell=8$, but here we explore the multipolar
dependence, truncating the sum at $\ell = 2$ (circles), $\ell = 3$
(squares), $\ell = 4$ (diamonds), $\ell = 5$ (triangles up), $\ell =
6$ (triangles down) and $\ell = 8$ (star).
Observe that for small velocities (i.~e.~$v = 0.01$ or $v=0.1$) the
remainder generally decreases with PN order when using up to $\ell = 8$
harmonics.  This is precisely the convergent-like behavior alluded to
earlier. Also observe that an increasing number of multipole moments
must be included as a function of PN order to perform any type of
meaningful comparison between 
PN theory and numerical results. We shall study how this number of
multipoles depends on velocity in Sec.~\ref{mult-exp}.

\section{Region of validity}
\label{reg-of-val}

Before applying asymptotic techniques to determine the region of validity, let
us show how a quick but {\emph{incorrect}} estimate can be made directly through
Fig.~\ref{remainder-PN-EMR}. Let us then forget for the moment that
higher-order approximations should be more accurate than lower-order ones, and
simply draw a horizontal line in Fig.~\ref{remainder-PN-EMR} at some maximum
error threshold, such as $\delta_{\rm{max}} = 0.001$. The required accuracy,
of course, depends on the application one has in mind.  One can then guarantee
that inside some region $v < \tilde{v}$ the error in the $N$th-order PN
expansion is less than $\delta_{\rm{max}}$. 

\begin{table}[htb]
\begin{center}
\begin{tabular}{c|c|c|c|c|c|c|c|c|c|c}
\hline
\hline
$N$ & $2$ & $3$ & $4$ & $5$ & $6$ & $7$ & $8$ & $9$ & $10$ & $11$ \\
\hline
$\tilde{v}$ & $0.04$ & $0.11$ & $0.14$ & $0.14$ & $0.19$ & $0.30$ &
$0.22$ & $0.25$ & $0.28$ & $0.29$ \\
\hline
$\delta F^{(N)} [$\%$]$ & $0.10$ & $0.10$ & $0.10$ & $0.10$ & $0.11$ & $0.11$ & $0.11$ & $0.11$
& $0.11$ & $0.11$\\
\hline
\hline
\end{tabular}
\end{center}
\caption{\label{table0} For $v < \tilde{v}$ (second row) the error of using
  the $N$th order PN expansion (where $N$ is listed in the first row) is less
  than $0.001$, i.~e.~ the relative fractional error $\delta
  F^{(N)} \approx 0.1\%$ (second row).}
\end{table}
We present $\tilde{v}$ in Table~\ref{table0}, together with the relative
fractional error in the energy flux, $\delta F^{(N)} = (F-F^{(N)})/F$. This
table shows, as expected, that the fractional relative error in the flux is
approximately always of the same order, $0.1\%$. This scheme, however, forces
the $11$th-order PN approximant to be as accurate as the $2$nd-order PN
approximant, which is clearly inconsistent with perturbation theory.

The restriction that higher-order approximations be more accurate than
lower-order ones is naturally incorporated in the asymptotic estimates
of the region of validity, which we shall perform next. The condition
defining the edge of the region of validity is
\be
\label{E-order}
{\cal{O}}(F - F^{(N)}) = {\cal{O}}(F^{(N+1)} -
F^{(N)})\,. 
\ee
In other words, we have to determine the $\bar{v}$ for which the
remainder ceases to be comparable to the next order term.

\begin{figure}[htb]
  \epsfig{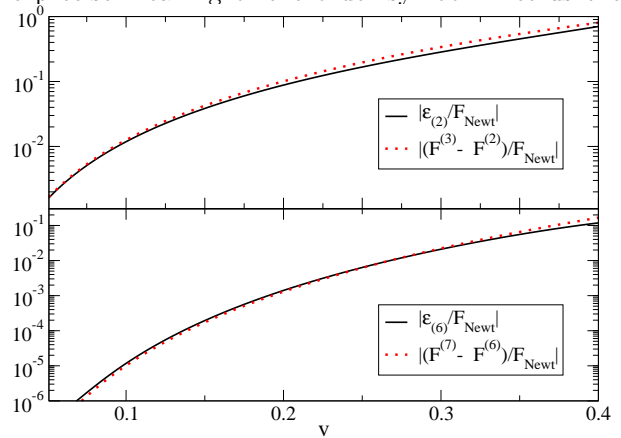}
\caption{\label{region-PN-EMRI} Top: Log-linear plot of the absolute value of
  the remainder of the $N=2$ PN flux in the extreme-mass ratio limit (solid)
  and the $N=3$ term (dotted).  Bottom: Same as top, but for the $N=6$
  remainder and the $N=7$ term.  All curves have been factored by the
  leading-order, Newtonian expression for the energy flux.}
\end{figure}

The remainder and the next-order term possess two distinct and generic
features when compared to each other. Figure~\ref{region-PN-EMRI}
presents the $n=2$ ($n=6$) order remainders, as well as the $n=3$
($n=7$) order term in the top (bottom) panel as a function of
velocity. The two distinct behaviors alluded to earlier are the
following: either the remainder and the next-order term are of the
same order for sufficiently low velocities, until eventually the
curves separate for larger velocities; or the next-order term starts
off smaller than the remainder, but eventually the curves cross and
separate. When the curves cross (in the bottom panel, this crossing
occurs roughly at $v \approx 0.275$) one can apply the same techniques
of the previous section and simply define $\bar{v}$ as the velocity
where the curves intersect. However, when the curves do not cross (top
panel), we must extend the methods we used for Bessel functions.

The asymptotic analysis definition of the edge of the region of
validity is inherently ambiguous, depending on the precise meaning of
the order symbol. Let us then replace Eq.~(\ref{E-order}) by
\be
\left|F - F^{(N)}\right| - \left|F^{(N+1)} -
F^{(N)}\right| < \delta\,,
\ee
where $\delta$ is some tolerance. One would expect this tolerance to
decrease with PN order, since higher-order approximations should be
more sensitive to smaller differences. Let us, however, forget this
for the moment and demand a {\emph{constant}} tolerance $\delta =
10^{-3}$. This procedure is somewhat similar to picking a maximum
error threshold in the remainders of the approximation, which we
already explored in Table~\ref{table0}. Here, however, we do not
demand a constant maximum error, but instead we arbitrarily choose a
constant {\emph{relative difference}} between the remainder and the
next term in the series. 

\begin{widetext}
\begin{center}
\begin{table}[htb]
\begin{tabular}{c|c|c|c|c|c|c|c|c|c}
\hline
\hline
$N$ & $2$ & $3$ & $4$ & $5$ & $6$ & $7$ & $8$ & $9$ & $10$\\
\hline
$\bar{v}(\delta\bar{v})$ & $0.107(0.017)$ & $0.138(0.021)$ & $0.140(0.017)$ & $0.190(0.020)$ & $0.266(0.045)$ & $0.222(0.019)$ & $0.248(0.019)$ & $0.281(0.018)$ & $0.292(0.019)$ \\
\hline
$\delta F [\%]$ & $1.51$ & $0.29$ & $0.11$ & $0.61$ & $1.03$ & $0.02$ & $0.28$ & $0.35$ & $0.16$ \\
\hline
\hline
\end{tabular}
\caption{\label{table2} Approximate values of the edge of the region
  of validity for different orders of the EMRI-PN energy flux. The first row
  lists $\bar{v}$ and (in parentheses) $\delta\bar{v}$; the second row lists
  $\delta F^{(N)} = (F - F^{(N)})/F$, evaluated at $\bar{v}$.}
\end{table}
\end{center}
\end{widetext}

Applying this procedure, one obtains the $\bar{v}$ and its error
$\delta\bar{v}$ presented in Table~\ref{table2}.  Here $\delta\bar{v}$
symbolizes variations of the tolerance in the interval $\delta = \{10^{-3}/5,5
\times 10^{-3}\}$, i.~e.~we evaluate $\bar{v}_1$ ($\bar{v}_2$) with $\delta_1
= 10^{-3}/5$ ($\delta_2 = 5 \times 10^{-3}$) and define the error as
$\delta\bar{v} := |\bar{v}_1 - \bar{v}_2|/2$. As one can see from
Table~\ref{table2}, the relative fractional error decreases on average with PN
order, although not monotonically.

\begin{figure}[htb]
\epsfig{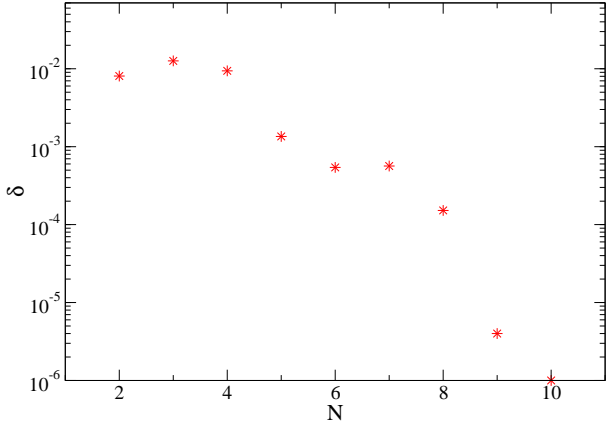}
\caption{\label{tolerance} Tolerance as a function of PN order.}
\end{figure}
Higher-order approximations should be sensitive to a smaller tolerance, which
implies that this quantity cannot be set arbitrarily.  Instead, $\delta$
should be given by the error in the difference between the $N$th remainder and
the $(N+1)$th-order term.  This error is presumably of the order of the error
in the $(N+1)$th-order term, and it can be estimated by the $(N+2)$th-order
term.  This quantity, of course, depends on $v$, but its order can be roughly
given by its absolute value evaluated in the middle of its range $v \approx
0.2$.  By this method we estimate that the tolerance behaves with PN order as
shown in Fig.~\ref{tolerance}.  Since this estimate of the tolerance is not
exact, we shall allow $\delta$ to vary between $5$ times and $1/5$ times its
value. In this way, we can determine the sensitivity of our results to the
choice of $\delta$ and provide error bars as done previously.

The edge of the region of validity can now be computed using the
tolerance criterion defined above.  Table~\ref{table3} presents
$\bar{v}$, together with an averaged error bar $\delta\bar{v}$, which
represents variations of $\{\delta/5.0,5.0 \; \delta\}$. The first
line in the first row of this table uses the tolerance presented in
Fig.~\ref{tolerance}.  Observe that the relative fractional error in
the flux, the first line in the second row of Table~\ref{table3}, decreases on average
as the order of the approximation increases. However, the region of
validity seems to increase between $2$nd and $6$th PN order (except as
one goes from $4$ to $5$), while it seems to decrease between $6$th
and $10$th PN order.  We shall analyze this behavior in more detail
shortly.

%
\begin{widetext}
\begin{center}
\begin{table}[htb]
\begin{tabular}{c|c|c|c|c|c|c|c|c|c}
\hline
$N$ & $2$ & $3$ & $4$ & $5$ & $6$ & $7$ & $8$ & $9$ & $10$ \\
\hline
\hline
$\bar{v}(\delta\bar{v})$ &$0.179(0.029)$  & $0.200(0.014)$  & $0.207(0.025)$  & $0.199(0.021)$  & $0.288(0.007)$  & $0.207(0.018)$ & $0.204(0.016)$  & $0.178(0.010)$  & $0.160(0.006)$  \\
&$0.160(0.026)$  & $0.200(0.014)$  & $0.214(0.026)$  & $0.197(0.021)$  & $0.346(0.019)$  & $0.214(0.018)$ & $0.207(0.016)$  & $0.158(0.005)$  & $0.179(0.029)$  \\
&$0.143(0.023)$  & $0.200(0.014)$  & $0.221(0.027)$  & $0.195(0.020)$  & $0.390(0.024)$  & $0.222(0.019)$ & $0.211(0.016)$  & $0.160(0.101)$  & $0.160(0.101)$  \\
\hline
$\delta F [\%]$ &$6.7607$  & $-1.3250$ & $-0.5562$ & $0.7888$  & $-1.6931$ & $-0.0134$ & $0.0518$  & $-0.0043$ & $0.0001$ \\
&$ 4.8505$ & $-1.3250$ & $-0.6269$ & $ 0.7531$ & $-5.2159$ & $-0.0151$ & $ 0.0592$ & $-0.0014$ & $ 0.0005$\\
&$ 3.5011$ & $-1.3250$ & $-0.7065$ & $ 0.7103$ & $-9.6392$ & $-0.0168$ & $ 0.0686$ & $-0.0015$ & $ 0.0001$\\
\hline\hline
\end{tabular}
\caption{\label{table3} First row: edge of the region of validity
  $\bar{v}$ and (in parentheses) estimated error $\delta\bar{v}$. Second
  row:  $\delta F^{(N)}$ [$\%$] with the tolerance set by the $(N+2)$th-order
  term in the approximation. Top to bottom, we list values corresponding to
  three successive iterations of our attempt to estimate the region of
  validity (see text).}
\end{table}
\end{center}
\end{widetext}
%

Before discussing the implications of these results, let us attempt to
determine how reliable they are by experimenting with the choice of $\delta$.
Let us then continue to define the tolerance through the $(N+2)$th-order term,
but this time let us evaluate it at the value of $\bar{v}$ found in the first
line of the first row of Table~\ref{table3}. Doing so, we obtain the values of $\bar{v}$
presented in the second line of the first row of this table. If we
iterate this algorithm once more and 
use the second line in the first row of Table~\ref{table3} to evaluate the $(N+2)$th-order
term, we obtain yet another $\bar{v}$, given in the third line of the
first row of this table.

The edge of the region of validity seems to be rather insensitive to the
choice of $\delta$, provided the $(N+2)$th-order term is not evaluated too far
from the mean of the domain. Indeed, Table~\ref{table3} shows that for most
values of $N$ the value of $\bar{v}$ given in the first row remains
within the error bars of the first line of the first row.  This is not the case for $N =
(9,10)$, because then the values of $\delta F$ obtained are at the level of
the numerical accuracy of our simulations: $\delta F^{(9,10)} \approx
10^{-6}$.  This is also not the case for $N = 6$, because this is an
inflection point in the behavior of the edge of the region of validity, and
thus the first iteration forces $\bar{v}$ outside of this region.  We
therefore conclude that the first line in the first row of Table~\ref{table3} suffices as a
reliable, approximate measure of the edge of the region of validity of the PN
approximation.

A note of warning is due at this point: the iterative scheme presented
here should not be confused with a convergence scheme. One might be
tempted to expect that as we continue to iterate, $\bar{v}$ will tend
to some definite value that will inequivocably define the region of
validity exactly. This concept, however, is inherently flawed because
asymptotic series by definition do not possess an exact region of
validity. As already explained, asymptotic series and their region of
validity are defined via order symbols, and thus can only be
interpreted as approximate concepts.

\begin{figure}[htb]
\epsfig{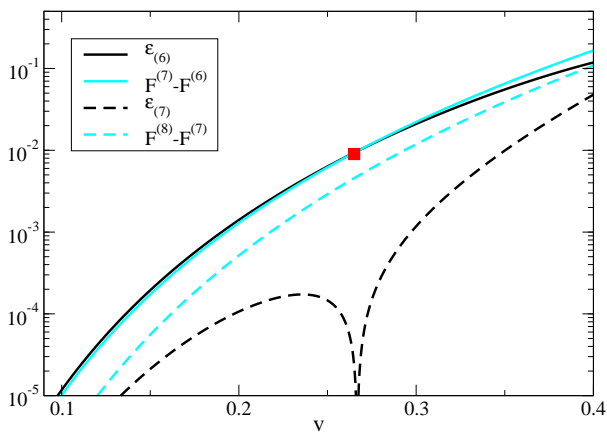} 
\caption{\label{weird} Plot of $\epsilon_{(6)}$ (solid black),
  $\epsilon_{(7)}$ (dashed black), the $7$th-order term (solid light
  gray) and the $8$th-order term (dashed light gray). The square shows
  where $\epsilon_{(6)}$ and $(F^{(7)} - F^{(6)})$ cross.}
\end{figure}

Let us now discuss the behavior of the edge of the region of validity with PN
order in more detail. As we already mentioned, there seems to be an inflection
point at $N=6$: on average, $\bar{v}$ increases for $N<6$, while it decreases
for $N>6$. Such a behavior then forces the $N=6$ approximation to have the
largest region of validity. By naively inspecting Fig.~\ref{remainder-PN-EMR}
one could be tempted to conclude that this result is paradoxical, since
$\epsilon_{(7)} < \epsilon_{(6,8)}$. However, recall that the region of
validity is determined by the difference between $\epsilon_{(N)}$ and the
$(N+1)$th-order term ({\em not} $\epsilon_{(N+1)}$).  To stress this
difference, in Fig.~\ref{weird} we show $\epsilon_{(6)}$, $\epsilon_{(7)}$
(black) and the $7$th- and $8$th-order terms (light gray).  Although the
absolute value of $\epsilon_{(7)}$ is smaller than $\epsilon_{(6)}$, observe
that $\epsilon_{(6)} \approx F^{(7)} - F^{(6)}$ in a large velocity range,
until the curves cross around $\bar{v}= 0.265$. On the other hand,
$\epsilon_{(7)} \approx F^{(8)} - F^{(7)}$ only for $v < 0.2$, and soon after
the curves separate. We see then that the absolute magnitude of the remainder
itself does not determine the region of validity.  As we stress once more,
this region is defined by the requirement that the error in the approximation
is smaller than (or of the same order as) the next-order term in the series.

The behavior of the edge of the region of validity with PN order was already
presented in the Introduction (Fig.~\ref{intro:regions-final-PN-EMR}). Whether
there is truly a negative slope for $N>9$ is difficult to establish due to the
truncation error, but our results seem to support this hypothesis.
Nonetheless, we do observe in the figure that for $N<6$ the region of validity
increases (as expected of a convergent series), while for $N>6$ it decreases
(as expected of a divergent series). Such a result seems to agree with the
statement that the logarithmic terms in the PN approximation somewhat destroy
the convergence properties of the
series~\cite{Porter:2007vk,Porter:2006cn,Porter:2005cu}.  Nonetheless, as we
have seen in this paper, a divergent series can sometimes be even more useful
and accurate than a convergent one, provided it is used within its region of
validity.

\begin{figure}[htb]
\epsfig{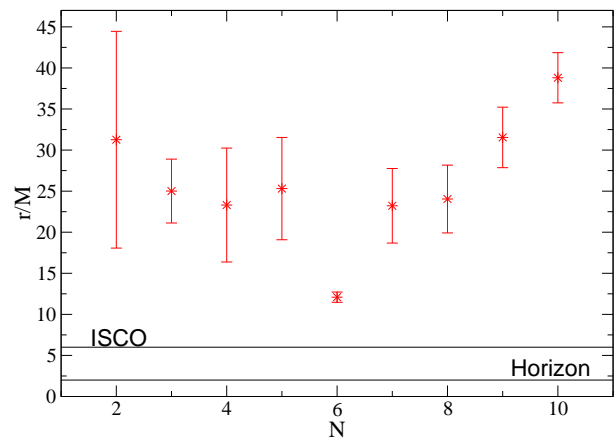}
\caption{\label{regions-final-PN-EMR-distance} Minimum orbital radius
  [Eq.~(\ref{radius})] representation of the edge of the region of
  validity for different PN orders.}
\end{figure}
The edge of the region of validity can also be presented as a function of the
Schwarzschild radius of the particle's circular orbit. Figure
~\ref{regions-final-PN-EMR-distance} shows the minimum Schwarzschild radius
delimiting the region of validity, as a function of PN order. This radius can
be obtained from $\bar{v}$ through the generalization of Kepler's law to
circular geodesics:
\be\label{radius}
\frac{r}{M} = \frac{1}{v^2}\,.
\ee
Simone {\emph{et al.~}} determined that the PN approximation for the infall of
a particle into a Schwarzschild black hole can be trusted down to a harmonic
radius $r_{\rm harm}\simeq 10M$ \cite{Simone:1995qu}. Perhaps it is no
coincidence that their final result is surprisingly close to the edge of the
region of validity for the 3PN approximant of the flux for quasi-circular
inspirals.

The results presented so far should not be misinterpreted as saying, for
example, that the 3PN approximation is more accurate than higher-order ones.
In fact, the statements made here say nothing about the absolute accuracy of
the PN approximation. These results only suggest relational statements between
the $N$th and the $(N+1)$th-order approximations. Table~\ref{table3} should be
read as saying that, for velocities $v < \bar{v}$, the $N$th order
approximation has errors that are of expected relative size. For larger
velocities, the $(N+1)$th- and higher-order terms become important, and should
not be formally neglected.  However, if one is willing to tolerate errors
larger than those estimated by the $(N+1)$th order term, and if one is willing
to relinquish the desire that higher-order approximation be more accurate than
lower-order ones, then one can surely go beyond $\bar{v}$ at the cost of
losing analytic control of the magnitude of the error.

\section{Multipoles and PN orders}
\label{mult-exp}

As a by-product of this analysis, we can ask the following interesting
question: how many multipoles should be kept in the numerical solution
if we are interested in studying the $N$th-order PN approximation? In
other words, we wish to determine whether it is sufficient to keep
only the quadrupole in certain cases, or if we always need higher
multipoles to study the accuracy of the PN series (see \cite{Mino:1997bx} for
a related discussion).

\begin{figure}[htb]
\epsfig{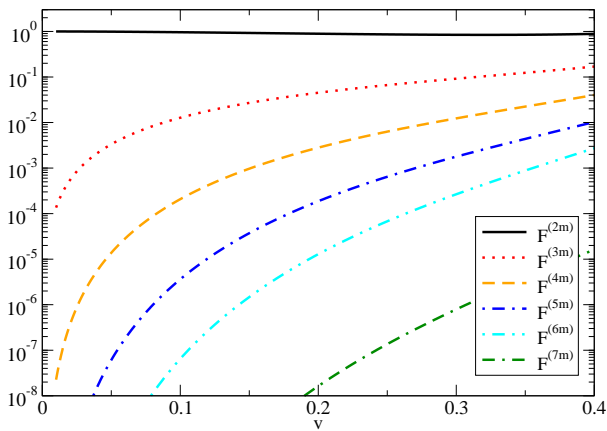}
\caption{\label{modes-PN-EMRI} Plot of the energy
  flux decomposed into spherical harmonics and summed over $m$. Each
  multipolar component is normalized to the Newtonian flux. The $\ell
  = 2$ harmonic is shown with a solid line, the $\ell = 3$ with a
  dotted line, the $\ell = 4$ with a dashed line, the $\ell = 5$ with
  a dot-dashed line, the $\ell = 6$ with a dot-dot-dashed line, and
  the $\ell = 7$ with a dash-dash-dotted line.}
\end{figure}

The answer to this question depends on the behavior of each multipolar
contribution to the energy flux as a function of velocity. In
Fig.~\ref{modes-PN-EMRI} we plot multipolar contributions with different
$\ell$'s (where for each $\ell$ we sum over all values of $m$) normalized to
the Newtonian flux.  The dominant $(\ell = 2)$ mode is very close to unity for
all values of $v$, while all other contributions go to zero as $v \to 0$ and
approach a roughly constant, $\ell$-dependent value. We do not show in the
figure the $\ell = 8$ harmonic to avoid cluttering, but this contribution
would be located at the lower right corner of the figure.

Figure~\ref{modes-PN-EMRI} implies that it is ludicrous to compare PN with
numerical relativity in certain systems if only a few multipoles are taken
into account.  For example, neglecting the $\ell = 4$ multipole leads to
errors of ${\cal{O}}(10)^{-4}$ at $v= 0.1$, which is comparable to terms of
${\cal{O}}(1/c)^4$. Thus, comparing PN expansions of ${\cal{O}}(1/c)^5$ and
smaller to numerical solutions neglecting the $\ell = 4$ multipole at $v <
0.1$ is risky and may lead to incorrect conclusions. The relative importance
of higher-$\ell$ multipoles decreases for equal mass systems, but again
recovers great importance when eccentricity and/or spins are included.

\begin{figure}[htb]
\epsfig{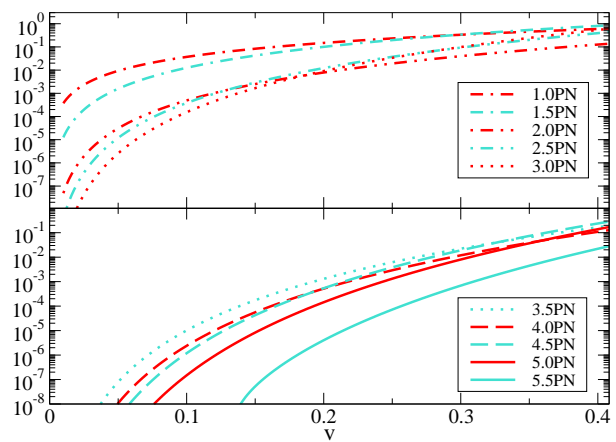}
\caption{\label{PN-contrib} Plot of the relative PN contributions to
  the total energy flux normalized to the Newtonian flux. Top: PN
  orders ${\cal{O}}(1/c^2)$ to ${\cal{O}}(1/c^6)$. Bottom: PN
  orders ${\cal{O}}(1/c^7)$ to ${\cal{O}}(1/c^{11})$.}
\end{figure}

The number of multipoles to retain also depends on the PN contribution
to the energy flux. Figure~\ref{PN-contrib} plots these contributions,
normalized to the Newtonian flux. 
As before, odd (even) orders are plotted in light (dark) gray. Observe that
the PN contributions of the top panel (lower PN orders) rise rather fast and
approach a roughly constant value. The opposite is true for the bottom panel,
where the curves are monotonically increasing in this velocity domain.

\begin{figure}
\epsfig{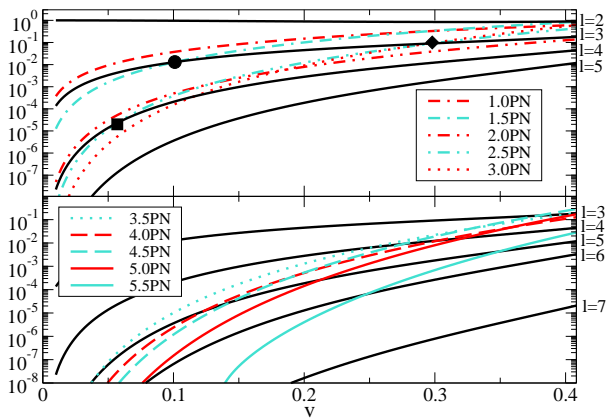}
\caption{\label{comparison} Same as Figs.~\ref{modes-PN-EMRI}
  and~\ref{PN-contrib}. The black solid curves are the harmonic
  contributions to the energy flux, while the non-solid curves
  represent the PN contributions. The intersections in this figure
  (some of which are shown as a square, a diamond or a circle) lead to
  Table~\ref{table:lmin} in the Introduction.}
\end{figure}

An analysis of the accuracy of the $N$th-order PN approximation
requires that the numerical error be at the very least smaller than
the $N$th-order term. In the asymptotic analysis of the previous
section, we simply summed up to $\ell = 8$ and always included all
harmonics in our analysis. However, it is possible that we did not
need to keep up to $\ell = 8$ in the analysis of all PN orders.
Figure~\ref{comparison} superimposes the PN contribution and the
multipolar contribution of the energy flux.
We see, for example, that the $\ell = 5$ harmonic contribution is
always negligible if one is studying a 1.5PN order accurate expression. 

The intersection of the harmonic and PN contributions provide a
{\emph{minimum}} requirement of accuracy if any type of comparison between
numerical simulations and PN theory is to be carried out.  These intersections
form the basis of Table~\ref{table:lmin} in the Introduction. For example, if
one wishes to compare numerical results to the $1.5$PN energy flux, then one
can simply use the $\ell = 2$ harmonic provided $v > 0.101$ (shown as a circle
in the figure). For smaller velocities, however, the $\ell = 3$ harmonic
contribution needs to be included. Similarly, if one is studying the 2.5PN
expression of the energy flux, then one can simply use the $\ell = 2$ harmonic
if $v>0.298$ (shown as a diamond in the figure), but one must include the
$\ell = 3$ harmonic if $0.057<v<0.298$ and the $\ell = 4$ harmonic if $v <
0.057$ (a square in the figure). Finally, note that the number of multipoles
that need to be included is larger for smaller velocities. This is because the
multipolar contribution raises steeply for small velocities and then roughly
asymptotes to a constant, leading to more intersections when $v$ is small.

\section{Conclusions and outlook}
\label{conclusions}

We have proposed a generic, gauge-independent and systematic method to
determine the formal region of validity of the PN approximation. This
method relies heavily on tools from asymptotic analysis, and in
particular, on the concept of an optimal asymptotic expansion. The
main philosophy of the approach is to determine the velocity or
frequency beyond which the next-order term in the series must be
included. This velocity is found by studying the region where the true
error in the approximation (relative to the exact or numerical answer)
becomes comparable to the series truncation error (due to neglecting
higher-order terms in the series).

We applied this method to the quasi-circular inspiral of a small compact
object into a non-spinning black hole. The PN approximation is thus coupled to
an EMR expansion, leading to what we here called the PN-EMR approximation. The
exact solution is modeled by a numerical calculation in black hole
perturbation theory. This scheme linearizes the Einstein equations in the
metric perturbation but not in the velocity parameter, and so it is suitable
for the study of EMR systems. Here we are mainly interested in a
proof-of-principle of the proposed method and we used the energy flux as the
dependent variable, but we could have studied other time-dependent quantities.
The gravitational wave phase, for example, is the most interesting quantity
for data analysis purposes. However, relating the energy flux to the phasing
involves a certain degree of arbitrariness (see \cite{Boyle:2007ft} for a
discussion) and we preferred to avoid these complications in our preliminary
exploration.

The output of the method is the edge of the region of validity as a
function of PN order (Fig.~\ref{intro:regions-final-PN-EMR}). We found
that the 3PN approximation of the energy flux seems to possess the
largest region of validity.  This implies that the 3PN approximation
can be evaluated up to rather high velocities (i.~e.~$v \approx 0.29$)
and still produce an approximate answer with an error of expected size
(given by the next order term in the series).  Moreover, we found that
for lower than 3PN order, the PN approximation seems to present a
convergent-like behavior (the region of validity increases with PN
order).  Similarly, PN approximants of order higher than 3PN present a
divergent-like behavior (the region of validity decreases with PN
order), probably associated with the appearance of logarithmic terms
in the series.

Another result of this paper is the {\emph{minimum}} number of
multipoles that need to be retained to perform any type of comparison
with PN theory (Table~\ref{table:lmin}). We find that the number of
multipoles depends on the velocity regime where the comparison takes
place. In particular, we find that more multipoles are required for
low velocities.  At $3.5$PN order, the inclusion of the first five
multipoles suffices to cover the entire velocity range. It will be
interesting to explore the number of multipolar components that must
be included in a numerical simulation of comparable mass binaries to
get some required accuracy in the {\em phasing} of the waves.

Although the method is generic and gauge-independent, the conclusions derived
from the analysis of EMR systems are not necessarily generalizable to
comparable mass systems. The number of multipoles needed to match the flux in
the comparable mass case is probably smaller than suggested here. Moreover,
the PN series is suspected to be much less accurate for EMR systems, which
suggests that perhaps for comparable mass systems the onset of the
divergent-like behavior observed here could occur at higher PN orders.

Various extensions of this work should be possible in the near future. In
particular, we would like to consider: 
\begin{enumerate}
\item[1)] Other physical systems: Although we have here studied only
  non-spinning EMR systems, the proposed method is very general and it can be
  applied to more complex physical scenarios. For example, by modeling a small
  particle in orbit around a spinning (Kerr) black hole, one could study
  whether the black hole's angular momentum substantial affects the region of
  validity of PN theory. Other interesting scenarios would be comparable-mass,
  eccentric and spinning binaries. These systems, however, are much more
  difficult to model, both numerically and in PN theory.  In particular, an
  accurate estimate of numerical errors in present numerical relativistic
  simulations is quite
  challenging~\cite{Buonanno:2006ui,Berti:2007fi,Boyle:2007ft}.
  
\item[2)] Different observables: We have here considered only the
  energy flux, because it is a well-defined, time-dependent
  observable. We could, however, also apply this method to the phasing
  of gravitational waves or the linear and angular momentum fluxes.
  One expects the asymptotic properties of PN theory to be independent
  of the quantity analyzed to determine it. If this were the case, the
  regions of validity found by analyzing, for example, the momentum
  flux or the phasing should be comparable to those found here.
  Although unlikely, if this were not the case, then different
  PN-expanded quantities of the {\emph{same}} order would possess
  different regions of validity, casting doubts on the asymptotic
  structure and consistency of the PN approximation.
 
\item[3)] Different PN flavors: The method considered here does not
  depend on the PN approximant we use. We would like to study the
  region of validity of other (non-Taylor) PN approximants, such as
  the effective-one-body
  approach~\cite{Buonanno:1998gg,Buonanno:2000ef},
  Pad\'e~\cite{Damour:1997ub,Damour:2000gg} and Chebyshev
  resummations~\cite{Porter:2005cu,Porter:2006cn,Porter:2007vk}.  In
  the EMR case, a fully convergent expansion of the flux might in
  principle be obtained via the Mano-Takasugi
  method~\cite{Mano:1996gn,Mano:1996mf,Mano:1996vt,Fujita:2004rb}.  To
  our knowledge, however, the adiabatic energy flux has not been
  worked out explicitly in this approach, and their method does not
  seem to be straightforwardly generalizable to comparable masses.

\end{enumerate}

The studies suggested above would answer a number of questions and
shed light on the structure of the PN series. Point (1) would reveal
how the region of validity depends on the initial properties of the
system. Based on previous studies \cite{Blanchet:2002xy,Mora:2003wt},
one would expect this region to increase as the mass ratio approaches
unity.  Moreover, point (1) would shed light on the possible overlap
of different approximation schemes, such as PN theory, black hole
perturbation theory and the close-limit
approximation~\cite{Price:1994pm}, providing support to asymptotically
matched global
solutions~\cite{Yunes:2005nn,Yunes:2006iw,nicos-t-funcs}. Point (3)
would allow us to make relative statements about the regions of
validity of different PN flavors. One could then verify whether
different resummation techniques truly increase the region of validity
of the PN approximation (see e.~g.~\cite{Blanchet:2002xy} for
criticism of this idea). In any case, a detailed study of the analytic
structure of the PN approximation should greatly benefit our
understanding of the dynamics of inspiraling compact binaries.

\acknowledgments
We are grateful to Abhay Ashtekhar, Carl Bender, Luc Blanchet, Alessandra
Buonanno, Vitor Cardoso, Curt Cutler, Jos\'e Gonz\'alez, Yasushi Mino, Ben
Owen, Eric Poisson, Uli Sperhake and Clifford Will for useful comments and
discussions.
N.Y. acknowledges support from National Science Foundation award PHY
05-55628 and the support from the Center for Gravitational Wave
Physics, which is funded by the National Science Foundation under
Cooperative Agreement PHY 01-14375. E.B.'s research was supported by
an appointment to the NASA Postdoctoral Program at the Jet Propulsion
Laboratory, California Institute of Technology, administered by Oak
Ridge Associated Universities through a contract with NASA. Copyright
2008 California Institute of Technology. Government sponsorship
acknowledged.

\clearpage
\appendix

\section*{Appendix: Logarithmic terms}
\label{app:logterms}

The presence of logarithmic terms in the PN expansion [see
e.~g.~Eq.~\eqref{E-EMRI-PN}] seems inherent to the PN approximation in
harmonic coordinates. The question we wish to answer is whether these
terms affect the use of standard asymptotic techniques. The answer to
this question is in the negative, and we explain why below.

First, Arun et al.~~\cite{Arun} have argued that the logarithmic terms can be
removed by a coordinate transformation for eccentric comparable-mass binaries.
The new, so-called modified harmonic coordinates differ from standard harmonic
coordinates to 1PN order and are regular everywhere.  Therefore, since the
convergence or divergence properties of the approximation should not depend on
the coordinate system used, it should not matter whether we do the analysis in
harmonic or modified harmonic coordinates. In other words, if we were to
repeat the analysis in modified harmonic coordinates, we should arrive at
similar conclusions as those found here.

Second, we can re-write the logarithmic terms in the PN approximation in a
manner which clearly exhibits its asymptotic structure. Let us define $\nu
\equiv \ln v$, such that $\nu \in [-\infty,0]$ as $v \in [0,1]$. Then we can
write
\be
\sum_{p=6}^{\infty} a_p v \ln(v^p) = e^{\nu} \sum_{p=6}^{\infty} b_p \nu\,, 
\ee
where we have absorbed a factor of $e^{p}$ into $b_p$. We then see that in the
new variables this series is reminiscent of a standard asymptotic expansion
with an exponential controlling factor (see e.~g.~\cite{Bender}).

Third, we can straightforwardly show that the series is indeed
asymptotic in some small velocity region. Let us first note that any
series $w_0 = \sum_n a_n v^n$ is indeed an asymptotic series for small
$v$, which is simply a restatement that any Taylor expansion is an
asymptotic series~\cite{Bender}.  Similarly, it follows that a series
of the form $w_1 = \sum_n b_n \ln(v)$ is {\em not} an asymptotic
series, because $\ln(v)$ diverges as $v\to 0$. Fortunately, the PN
expansion does not contain isolated logarithmic contributions.  All
logarithmic terms are multiplied by some power of the velocity,
i.~e.~the expansion has the form $w_2 = \sum_n b_n v^n \ln(v)$. As
$v\to 0$, such a series indeed tends to zero and is well-behaved.

Not only should the series be well-behaved as $v\to 0$, but it should also
decay at the right rate, i.~e.~in such a way so that Eq.~(\ref{asymptdef})
holds.  Let us then look at the $N=6$ term in the PN expansion of the energy
flux. Since we know that $a_6 v^6$ decays at the right rate, we can compare
$b_6 v^6 \ln(v)$ to this quantity.  One can easily verify that $a_6 v^6$ is
larger than the logarithmic term for $v \gtrsim 0.033$, while the opposite is
true for smaller velocities. When $v \approx 10^{-4}$ and smaller, the
logarithmic term dominates over the polynomial term, but such small velocities
are excluded from our analysis anyway, since we only compute the numerical
flux down to $v=10^{-2}$. Therefore the $N=6$ term in the PN expansion of the
energy flux is not only well-behaved as $v \to 0$, but it indeed decays at the
right rate in the velocity region considered here.

Based on these arguments, we conclude that there exists a
small-velocity {\emph{near zone}} region, where logarithmic terms in
the PN series do not affect the asymptotic analysis described in this
paper.

\section*{Erratum}

This erratum corrects some mistakes in Ref.~\cite{Yunes:2008tw}.

By comparison with high-accuracy numerical data produced with an independent
code by Scott Hughes, we found out that our numerical solution of the
Teukolsky equation (i.e., the numerical data used to approximate the exact
solution to the flux function) is only accurate up to $\ell = 6$. This implies
that the line labeled $F^{(7m)}$ in Fig.~\ref{modes-PN-EMRI}
of~\cite{Yunes:2008tw} is in error. For the same reason the data marked by
asterisks in Fig.~\ref{remainders-fix-v} are not accurate, and should be
ignored. Note also that the $x$-axis in Fig.~\ref{remainders-fix-v} should
range from $N=2$ to $N=11$: there is no PN correction to the flux when $N=1$.

More importantly, we found two mistakes in the code that computes the optimal
velocity of expansion (the edge of the region of validity). Fortunately, these
mistakes do not affect the main conclusion of~\cite{Yunes:2008tw}: that the
edge of the region of validity shrinks for $N>6$.  Below we discuss these
mistakes and correct them, providing updated tables and figures.

The first mistake affects the $N=3$ and $N=6$ data points.  In Fig.~$1$ and in
Table V of~\cite{Yunes:2008tw}, the edge of the region of validity was found
to be $\bar{v} = 0.2$ and $\bar{v}=0.29$ for the $N=3$ and $N=6$ cases,
respectively. These values were obtained through Eq.~$(20)$
in~\cite{Yunes:2008tw}:
\be\label{deltaN}
\delta_N(v)\equiv\left| |F-F^{(N)}|-|F^{(N+1)}-F^{(N)}| \right|<\delta\,.  
\ee
We will refer to this equation as the $\delta_{N}$ criterion. Upon more
detailed analysis, we have found that in these cases $\delta_N(v)$ presents
zero crossing, so $N=3$ and $N=6$ correspond to the first class of cases
discussed after Fig.~$8$ in~\cite{Yunes:2008tw}. When such zero-crossings are
present, they should be used to define the edge of the region of validity,
instead of the $\delta_{N}$ criterion. The correct edges of the region of
validity are $\bar{v} = 0.38$ and $\bar{v}=0.27$ for the $N=3$ and $N=6$
cases, respectively.

The second mistake affects the size of the error bars. In~\cite{Yunes:2008tw},
these were said to be given by solving Eq.~$(20)$ with $\delta \to \delta/5$
and $\delta \to 5 \delta$. This was a typo in the text: our code solved
Eq.~$(20)$ with $\delta \to \delta/2$ and $\delta \to 2 \delta$ instead,
yielding smaller error bars.
%
%

Moreover, for the $N=3$ and $N=6$ cases, the $\delta_{N}$ criterion to define
error bars cannot be applied. Instead, one can compute the error as the
difference between $\bar{v}$ and the velocity at which
$|F-F^{(N)}|=|F^{(N+2)}-F^{(N+1)}|$.  Doing so, however, would yield a very
large error bar ($\delta \bar{v} \approx 0.18$) for the $N=3$ case.
Alternatively, in these cases w can compute the error as the difference
between $\bar{v}$ and the {\em local maxima} in $\delta_N(v)$. This is the
approach we take here, and it will be justified in more detail in a follow-up
paper. Tables~\ref{table2-corr} and~\ref{table3-corr} are a corrected version
of Tables IV and V in~\cite{Yunes:2008tw}.  They assume a constant $\delta =
10^{-3}$ and the $\delta_{N}$ values in Fig.~$9$ of~\cite{Yunes:2008tw},
respectively. The error bars quoted for the $N=3$ and $N=6$ cases correspond
to the ``local maxima'' criterion described above. The primary change in these
plots is in the error bars and in the values of $\bar{v}$ for $N=3$ and $N=6$.
\begin{widetext}
\begin{center}
\begin{table}[htb]
\begin{tabular}{c|c|c|c|c|c|c|c|c|c}
\hline
\hline
$N$ & $2$ & $3$ & $4$ & $5$ & $6$ & $7$ & $8$ & $9$ & $10$\\
\hline
$\bar{v}(\delta\bar{v})$ & $0.11(0.04)$ & $0.14(0.05)$ & $0.14(0.04)$ & $0.19(0.04)$ & $0.30(0.03)$ & $0.22(0.04)$ & $0.25(0.04)$ & $0.28(0.04)$ & $0.29(0.04)$ \\
\hline
$\delta F [\%]$ & $1.5$ & $-0.30$ & $-0.1$ & $0.61$ & $-2.1$ & $-0.02$ & $0.28$ & $-0.35$ & $0.16$ \\
\hline
\hline
\end{tabular}
\caption{\label{table2-corr} Approximate values of the edge of the region of
  validity for different orders of the PN energy flux in the extreme
  mass-ratio limit. The first row lists $\bar{v}$ and (in parentheses)
  $\delta\bar{v}$; the second row lists $\delta F^{(N)} = (F - F^{(N)})/F$,
  evaluated at $\bar{v}$.}
\end{table}
\end{center}
\begin{center}
\begin{table}[htb]
\begin{tabular}{c|c|c|c|c|c|c|c|c|c}
\hline
$N$ & $2$ & $3$ & $4$ & $5$ & $6$ & $7$ & $8$ & $9$ & $10$ \\
\hline
\hline
$\bar{v}(\delta\bar{v})$ &$0.18(0.08)$  & $0.38(0.06)$  & $0.21(0.06)$  & $0.20(0.05)$  & $0.27(0.03)$  & $0.21(0.04)$ & $0.20(0.03)$  & $0.18(0.02)$  & $0.16(0.01)$  \\
&				          $0.16(0.07)$  & $\cdot$  	      & $0.21(0.06)$  & $0.20(0.05)$  & $\cdot$  & $0.21(0.04)$ & $0.21(0.03)$  & $0.16(0.01)$  & $0.17(0.04)$  \\
&					  $0.14(0.06)$  & $\cdot$  	      & $0.22(0.06)$  & $0.19(0.05)$  & $\cdot$  & $0.22(0.04)$ & $0.21(0.03)$  & $0.15(0.01)$  & $0.15(0.01)$  \\
\hline
$\delta F [\%]$ &$6.8$  & $-9.5$  & $-0.56$ & $0.79$  & $-1.0$ & $-0.013$ & $0.052$  & $-0.0043$ & $0.0001$ \\
&			     $4.8$ & $\cdot$ & $-0.63$ & $ 0.74$ & $\cdot$ & $-0.016$ & $ 0.060$ & $-0.0014$ & $ 0.00002$\\
&			     $3.5$ & $\cdot$ & $-0.71$ & $ 0.7$   & $\cdot$ & $-0.017$ & $ 0.070$ & $-0.0007$ & $ 0.00002$\\
\hline\hline
\end{tabular}
\caption{\label{table3-corr} First row: edge of the region of validity $\bar{v}$
  and (in parentheses) estimated error $\delta\bar{v}$. Second row: $\delta
  F^{(N)}$ [$\%$] with the tolerance set by the $(N+2)$th-order term in the
  approximation. Top to bottom, we list values corresponding to three
  successive iterations of our attempt to estimate the region of validity (see
  text). Error bars for the $N=3$ and $N=6$ terms are given by the local
  maxima criteria.}
\end{table}
\end{center}
\end{widetext}

\begin{figure*}[htb]
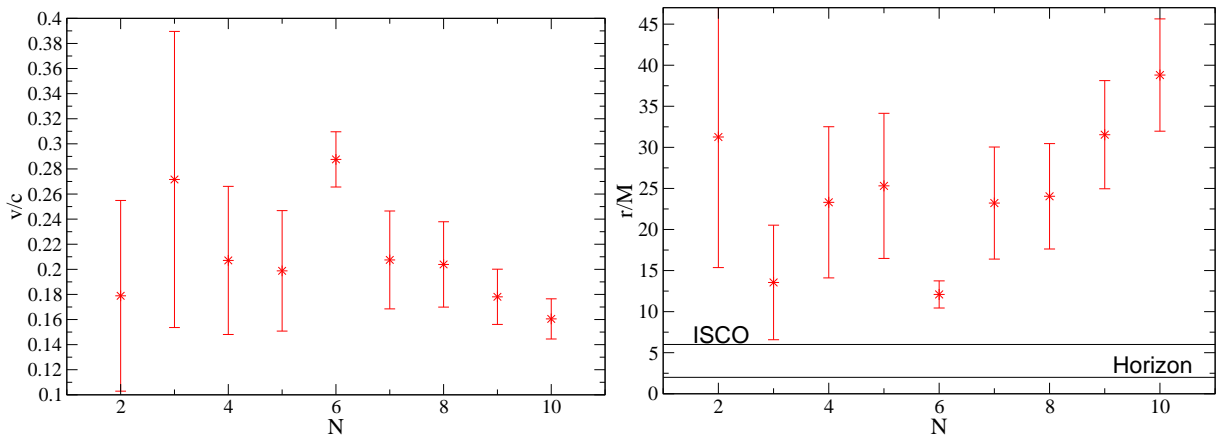

\epsfig{file=regions-final-PN-EMR-corrected.eps,width=8cm,angle=0,clip=true} 
\epsfig{file=regions-final-distance-PN-EMR-corrected.eps,width=8cm,angle=0,clip=true}
\caption{\label{intro:regions-final-PN-EMR-corr} Left: Edge of the region of
  validity for different PN orders. Right: Minimum orbital radius
  representation of the edge of the region of validity for different PN
  orders.  Error bars for the $N=3$ and $N=6$ terms are given by the local
  maxima criteria.}
\end{figure*}

The graphical representation of these tables must also be corrected. The
correct version of Figs.~$(1)$ and $(11)$ of~\cite{Yunes:2008tw} is
Fig.~\ref{intro:regions-final-PN-EMR-corr} of this Erratum. Observe that the
error bars have become larger. Moreover, notice that for $N=3$ and $N=6$
$\bar{v}$ is now much larger than for other values of $N$ (and the opposite
happens for $b$).  This is again because of the zero-crossing behavior of
these special points. More details will be given in an upcoming publication
\cite{paper2}. We reiterate, however, that although the $N=3$ and $N=6$ points
have changed, and the error bars have become larger, the main conclusion
of~\cite{Yunes:2008tw} (that the edge of the region of validity shrinks for
$N>6$) still holds.

Finally, note that Eq.~(18) of \cite{Yunes:2008tw} contains a typo: $M^2$
should be replaced by $M^{-2}$.

%


\end{document}